\documentclass[12pt]{article}

\usepackage{scicite}
\usepackage{graphicx}
\usepackage{braket}
\usepackage{amsmath}
\usepackage{color}
\usepackage{times}

\usepackage{graphicx}           % Include figure files
\usepackage{epstopdf}
\usepackage{subfigure}
\usepackage{graphicx}
\DeclareGraphicsExtensions{.eps,.ps}
\usepackage{pstricks}           % e.g. colors can be used in the text
\usepackage{dcolumn}            % Align table columns on decimal point
\usepackage{bm}                 % bold math
\usepackage{url}

\newenvironment{sciabstract}{%
\begin{quote} \bf}
{\end{quote}}

\newcounter{lastnote}

\title{\vspace{-3cm}Decoherence and Revival in Attosecond Charge Migration Driven by Non-adiabatic Dynamics}

\author
{Danylo T. Matselyukh$^1$, Victor Despr\'e$^2$, Nikolay V. Golubev$^3$, \\
Alexander I. Kuleff$^{2,\ast}$, Hans Jakob W{\"o}rner$^{1,\ast}$\\
 \\
\normalsize{$^{1}$ Laboratorium f\"{u}r Physikalische Chemie, ETH Z\"{u}rich, 8093 Z\"{u}rich, Switzerland}\\
\normalsize{$^{2}$ Theoretische Chemie, Physikalisch-Chemisches Institut (PCI), Universit\"at Heidelberg,}\\
\normalsize{69120 Heidelberg, Germany}\\
\normalsize{$^{3}$ Laboratory of Theoretical Physical Chemistry, Institut des Sciences et Ing\'enierie Chimiques,}\\
\normalsize{EPF Lausanne, 1015 Lausanne, Switzerland}\\
\normalsize{$^\ast$ \textbf{Corresponding authors. E-mails: alexander.kuleff@pci.uni-heidelberg.de, hwoerner@ethz.ch}}\\
\\
}

\date{}

\begin{document}

\baselineskip24pt

\maketitle

% Place your abstract within the special {sciabstract} enviropnment.
% 120-125 words
% opening sentence sets the question to general reader. background of this study, results, and concluding sentence.

\begin{sciabstract}
Attosecond charge migration is a periodic evolution of the charge density of a molecule on a time scale defined by the energy intervals between the electronic states involved. Here, we report the observation of charge migration in neutral silane (SiH$_4$) in 690~as, its decoherence within 15~fs, and its revival after 40-50~fs using X-ray attosecond transient absorption spectroscopy. The migration of charge is observed as pairs of quantum beats with a characteristic spectral phase in the transient spectrum. The decay and revival of the degree of electronic coherence is found to be a result of both adiabatic and non-adiabatic dynamics in the populated Rydberg and valence states. The experimental results are supported by fully quantum-mechanical {\it ab-initio} calculations that include both electronic and nuclear dynamics. We find that conical intersections can mediate the transfer of electronic coherence from an initial superposition state to another one involving a different lower-lying state. Operating on neutral molecules, our methods define a general approach to the key phenomena underlying attochemistry.
\end{sciabstract}

%%%%%%%%%%%%%%%%% main text %%%%%%%%%%%%%%%%
\newpage

%Introduction
\section{Introduction}

Charge migration in molecules is a purely electronic process driven by a coherent superposition of electronic states, which can appear due to a broadband excitation \cite{eyring44a,remacle06a} or electron correlation \cite{cederbaum99a,Kuleff14,kraus18a}. The time scales of charge migration are therefore directly defined by the inverse  of the electronic states' energy separations \cite{eyring44a,woerner17a}.
Since the electronic states constituting a superposition state typically have differently shaped potential-energy surfaces (PES), charge migration has been predicted to decohere rapidly because nuclear wavepackets evolving on different electronic states modulate the vertical energy interval or lose overlap in internal-coordinate or momentum space \cite{Vacher15,arnold17a,despre18a,jia19a}. Decoherence of charge migration results in a fading of the periodic charge rearrangement, and thus, together with non-adiabatic population transfer, to a more permanent transfer of the charge \cite{luennemann08a,woerner17a}, which can lead to bond-specific dissociation \cite{remacle98a,lehr05a}. As such, decoherence and non-adiabatic coupling are the key links between charge migration, that is by definition periodic in the few-states limit, and charge transfer, which is central to the further evolution of the induced dynamics and its final outcome \cite{Vacher15,arnold17a,despre18a,jia19a}.

The experimental observation of molecular charge migration remains a formidable challenge. Previous experiments have reported the reconstruction of attosecond charge migration in ionized iodoacetylene from high-harmonic spectroscopy \cite{kraus15b} and the observation of $\sim$4~fs quasi-periodic dynamics in the photofragmentation yield of ionized phenylalanine \cite{calegari14a}. These early results have triggered considerable theoretical activity \cite{Despre15,Kai-Jun17,Jia17}, which has also studied the role of nuclear dynamics in charge migration \cite{vacher14a,Lara-Astiaso17,Sun17,vacher2017,despre18a}. 
%problems with state of the art
%Whereas high-harmonic spectroscopy offers exquisite temporal resolution, on the order of 100~as \cite{baker06a,smirnova09b,haessler10a}, the accessible temporal window is limited to the transit time of the longest electron trajectory, i.e. on the order of 2-4~fs for typical driving wavelengths \cite{kraus18a}. This aspect prevented the observation of the intriguing effects of nuclear dynamics on charge migration that were subsequently predicted in a pioneering study \cite{jia19a}. 
%In previous pump-probe experiments combining attosecond and few-cycle pulses, the temporal resolution was typically limited by the few-cycle pulse duration \cite{goulielmakis10a,calegari14a,kobayashi18a, kobayashi19a}, which has motivated the development of light transients \cite{wirth11a,hassan16a,moulet17a}. Sub-pulse-length temporal resolutions have so far mainly been demonstrated in interferometric experiments, such as streaking \cite{kienberger03a}, photoelectron attosecond interferometry \cite{paul01a} and related experiments (e.g. \cite{neidel13a}), or in controlled free-induction-decay experiments \cite{ott14a, Drescher2019}. 

Here, we describe the first observation of decoherence and revival of attosecond charge migration. We use the same carrier-envelope-phase-(CEP)-stable few-cycle laser pulse to both excite charge migration in a neutral molecule and generate a soft-X-ray (SXR) pulse that probes the dynamics through attosecond transient-absorption spectroscopy (ATAS, \cite{goulielmakis10a,timmers19,kobayashi19a}). The choice of strong-field excitation was motivated by its sub-cycle time resolution and high excitation fractions, as compared to single-photon excitation. Whereas previous experiments on coherent superpositions of electronic states in neutral molecules have remained limited to the femtosecond time scale \cite{kraus13b,walt17a} and charge-migration experiments have remained limited to molecular cations \cite{kraus15b,calegari14a}, the present approach opens a pathway to studying charge migration in neutral molecules \cite{Dutoi2011}, while achieving sub-optical-cycle temporal resolution, thereby generalizing previous work on ground-state molecules \cite{neidel13a} and doubly-excited states of helium \cite{ott14a}. 
We find that the CEP-stable few-cycle pump pulse creates an electronic coherence between a valence- and a Rydberg-excited state, corresponding to a 690-as radial charge migration in SiH$_4$. Due to the differences in the associated PES, this electronic coherence is lost in $\sim$15~fs as the vibrational wavepackets in different electronic states oscillate with different periods. The electronic coherence is found to revive after 40-50 fs, as the two nuclear wavepackets transiently recover position- and momentum-space overlap. Most interestingly, the electronic coherence is additionally found to partially transfer to a second pair of states, involving a dark valence and the same upper Rydberg-like state. Whereas the observed de- and recoherence of charge migration through electronically adiabatic vibrational dynamics have recently been predicted to occur in the iodoacetylene cation \cite{jia19a}, we are not aware of a previous discussion of the transfer of electronic coherence, although it is related to the creation of electronic coherence at conical intersections \cite{kowalewski15a,keefer20a}. Importantly, these results show that electronic coherence can not only revive after being suppressed due to nuclear motion, but can even be transferred to other electronic states through conical intersections.

\section{Results and Discussion}

\subsection{Experimental Methodology and Results}
Figure~1 presents the experimental scheme and ATAS results. A 5.2-fs CEP-stable laser pulse centered at 780~nm is used to excite silane (SiH$_4$) molecules in the gas phase. The induced dynamics are probed by the transient absorption of an isolated attosecond pulse (with an estimated sub-200-as duration) covering the silicon L$_{2,3}$-edge. The time delay between the two pulses is stabilized to $<$25~as using a dual laser- and white-light interferometer \cite{huppert15a}. Details on the experimental methods are given in the supplementary material (SM, Section S1).
\begin{figure}
    \centering
    \includegraphics[width=1.0\textwidth]{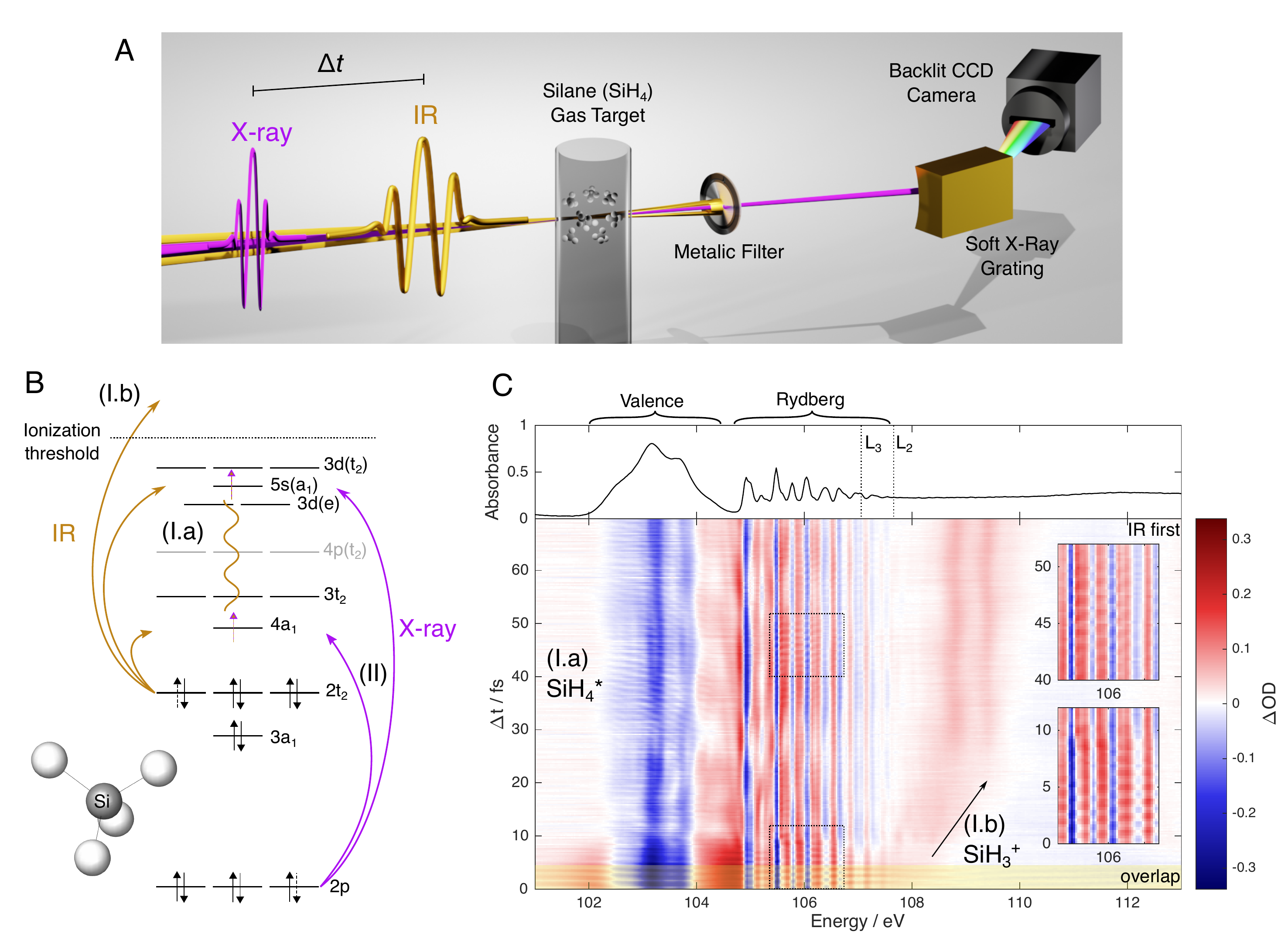}
    \caption{
    \textbf{Overview of experimental setup, pump-probe scheme and data.} A) Schematic representation of the experiment, B) molecular-orbital diagram of silane showing the ground-state configuration (black arrows), the preparation of an electronic superposition state by the pump pulse (yellow arrows), and its probing by core-level transient absorption (purple arrows). C) ATAS spectra as a function of time delay between the pump and probe pulses. The insets show a magnified view of the dotted boxes, highlighting the decay and revival of charge migration. The top panel shows the L$_{2,3}$-edge absorption spectrum of unexcited SiH$_4$.
    }
    \label{setupcluster}
\end{figure}
Figure~1B illustrates the excitation/probing scheme. The few-cycle pulse simultaneously excites (I.a) and ionizes (I.b) the sample. These two pathways are observed in different spectral regions, as illustrated in Fig.~1C. SiH$_4^+$ is unstable in its electronic ground state and therefore dissociates to form SiH$_3^+$+H. This fragmentation process occurs within 25~fs, manifesting as the appearance of a new absorption feature centred at 108~eV which then moves to higher energy to form two absorption bands between 108.7~eV and 109.4~eV, corresponding to SiH$_3^+$.

Henceforth, we concentrate on pathway I.a, i.e. strong-field excitation of valence and Rydberg states of the molecule
that lead to modulations of the absorption spectrum (top of Fig.~1C) between 102 and 107.5~eV. %Longer delay transient absorption measurements show that these states have a lifetime of up to 100~fs before fragmenting into SiH$_3$ and H (paper in preparation), giving ample time for coherences between these states to be studied. 
Our assignment of the static X-ray absorption spectrum of silane is based on core-valence-separated extended algebraic-diagrammatic construction through second order (CVS-ADC(2)-x) calculations shown as sticks in the top panel of Fig.~2A. The broad absorption band at 102-104~eV can be fit with four Gaussian functions, corresponding to the spin-orbit split excitations to 3t$_2$ and 4a$_1$ valence orbitals. The progression starting at 105~eV consists of d(e/t$_2$)- and s-Rydberg series including vibrational structure (not included in our stick spectra), converging to the L$_2$- and L$_3$-edges. All of our assignments agree with previous work \cite{puettner97a}, except for the relative ordering of the 3d(e) and 5s Rydberg states.

The most remarkable observation in Fig.~1C is the clear oscillation of the optical density (OD) in the spectral regions of 102-104~eV and 105.5-107.5~eV with a period of 1.31-1.39~fs that is observed to survive from the delays where pump and probe pulses overlap (yellow shading) out to about 15~fs (lower dashed box and right-hand inset), and to transiently revive between 40 and 50~fs (upper dashed box and inset). These rapid oscillations are a signature of charge migration \cite{golubev2020ATAS}.
%result of the interference of pairs of quantum paths upon the projection of a coherent superposition of electronic states into a common final state by the X-ray probe pulse. The pair of intermediate states leads to these quantum beats manifesting in ATAS as a pair of oscillating signals, separated in photon energy by the frequency of the quantum beat\cite{golubev2020ATAS}.

\subsection{Identification of Quantum Beats}

\begin{figure}
    \centering
    \includegraphics[width=0.8\textwidth]{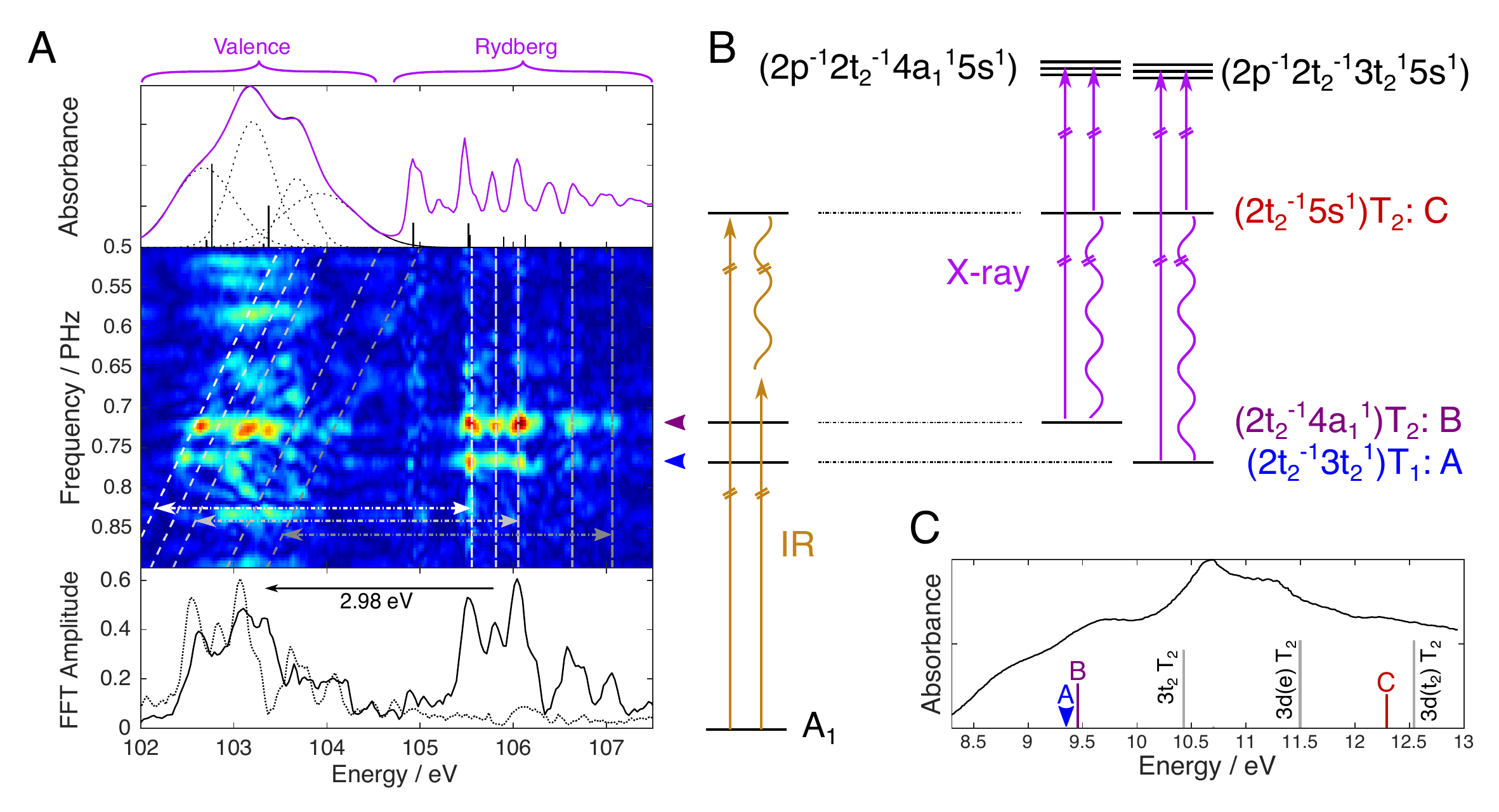}
    \caption{
    \textbf{Identifying the electronic states involved in charge migration.} A) Static absorption spectrum of silane at the Si L$_{2,3}$-edge with CVS-ADC(2)-x/aug-cc-pVTZ calculations (stick spectra shifted by -1.5 and -0.9~eV to account for the spin-orbit splitting of the L$_{2,3}$ edges) (top), spectrally-resolved FFT amplitude of the ATAS data (Fig.~1C) for delays above 10~fs (middle), line-out of the FFT amplitude at 0.72~PHz (bottom). B) Electronic-state diagram illustrating the preparation, time evolution, and probing of the electronic coherences. C) Valence absorption spectrum of silane (from \cite{Cooper95}) and calculated electronic absorption spectrum (sticks, EOM-CCSD/aug-cc-pVTZ). 
    %Shading represents the likelihood that a coherent buildup of the excited state population occurs throughout a few-cycle pulse depending on the final-state energy (for more details see section S2.6). 
    }
\end{figure}
The quantum beats are spectrally analyzed in the central panel of Fig.~2A, which presents a fast Fourier transform (FFT) along the delay axis (for $\Delta t > 10$~fs) of the ATAS results in Fig.~1C (for details, see Section S1.3). The FFT exhibits two distinct groups of frequencies at 0.720 and 0.765~PHz. Both consist of two similar sets of peaks separated by $\sim$3~eV.
In ATAS, electronic coherences are expected to appear as pairs of peaks fulfilling the relation $\Delta E=h/T=h\nu$, where $\Delta E$ is the energy separation of the electronic states, $T$ the period and $\nu$ the frequency of the quantum beat. Using the peaks in the Rydberg region as references (vertical white-to-grey shaded lines in Fig. 2A) and tracing the corresponding diagonal lines according to $\nu=\Delta E/h$, all strong FFT signals pair up. The line-out of the FFT amplitude at 0.72~PHz (bottom panel of Fig.~2A) demonstrates this more clearly. With the additional investigation of the phase of the quantum beats, discussed in section~\ref{Phase}, we are able to definitively assign these quantum beats to molecular charge migration.

The spectral position of the FFT signals with respect to the ground-state absorption spectrum of silane (top of Fig. 2A) also reveals the electronic configurations of the pump-prepared electronic states. 
The spectral overlap of all FFT signals in the Rydberg region indicates that the higher-lying state of both the 0.72-PHz and the 0.765-PHz coherence can only consist of an excitation into the 5a$_1$ orbital of 5s-Rydberg character. The observed spectral alignment is the consequence of the insensitivity of core-Rydberg transition energies to the valence electronic structure. 
By contrast, any states with 3d-Rydberg character would instead produce signals around 105.0~eV, which are not observed. The lower-lying states involved in the two coherences lie in the 102-104~eV spectral region, which uniquely identifies them as valence-excited states. Energetically, the 2t$_2\rightarrow$3t$_2$ and 2t$_2\rightarrow$4a$_1$ excitations are the only possible assignments, leading to the configurations given in Fig.~2B.

To relate these electronic configurations to specific valence-excited states, we performed equations-of-motion coupled-cluster singles-doubles (EOM-CCSD) calculations using an augmented correlation-consistent valence-triple-zeta (aug-cc-pVTZ) basis set (see SM, Section S2.1). The calculated electronic spectrum is compared to the measured valence-absorption spectrum of silane in Fig.~2C. The (2t$_2^{-1}$5a$_1^1$) and (2t$_2^{-1}$4a$_1^1$) configurations each give rise to a single state, with T$_2$ symmetry, henceforth designated as C and B states, respectively. The (2t$_2^{-1}$3t$_2^1$) configuration gives rise to a total of four electronic states of symmetries T$_1$, E, A$_1$ and T$_2$. Among those, the T$_1$ state is the only one to lie below the B state (see Fig. S5). %and therefore, given the measured 0.765-PHz frequency of the associated electronic coherence, represents the only possible lower-state assignment. 
Owing to its T$_1$ symmetry, this state labelled A, is dark in the single-photon absorption spectrum. The contribution of all these valence states to the observed quantum beats is determined with the help of fully quantum simulations.

\subsection{Fully Quantum Simulations of Electronic and Nuclear Dynamics}

Multiconfigurational time-dependent Hartree (MCTDH) calculations \cite{meyer1990multi,meyer2009multidimensional}  (for details see SM, Section S2) were performed on a manifold of 15 electronic states, spanning all four vibrational stretching modes of silane. PES of the lowest 16 states were calculated (Fig.~3A) at the EOM-CCSD/aug-cc-pVTZ level. These PES were then used to fit the parameters of a second-order vibronic-coupling Hamiltonian \cite{koppel1984multimode} that includes both non-adiabatic and Jahn-Teller interactions. These calculations provide a fully quantum-mechanical description of electronic and nuclear dynamics. Details of these calculations are discussed in the SM, section S2.2.

The strong-field nature of the pump pulse requires a careful discussion of the influence of the initial populations on the outcome of the MCTDH calculations. Experimental fluctuations of the pump pulse were considered by varying the initial phase of the states (for details see Section S2.3) and are represented by the shaded areas in Fig. 3B and 4B.  The less restrictive selection rules of multiphoton excitation compared to single-photon excitation were accounted for by studying two limiting cases. In the first case, all states contained in the MCTDH model were populated and in the second case, only the B and C states were initially populated. Details are given in the SM, Section S2.3. In both limiting cases the population in the C state remained constant over time. In the first case, the populations of the intermediate states were found to relax to the B and A states within 10~fs (Fig.~S8). In addition to the BC and AC coherences, these calculations predicted additional coherences at 0.65~PHz and 0.58~PHz (see Fig.~S8), which are not observed in the FFT shown in Fig.~2A. In the second limiting case, the population initially prepared in the B state, was found to relax to the A state in $\sim$7~fs (Fig.~3B). This B-A population transfer is mediated by conical intersections in the $\nu_4$ vibrational mode, shown in Fig.~3A (right) and visible in the dynamics of Movie S2. In agreement with the experiment (Fig.~2A), these calculations exclusively predict the appearance of BC and AC coherences. This comparison supports the predominant population of the B and C states by the pump pulse. The corresponding selectivity is consistent with few-cycle excitation in the strong-field regime, which results in spectral excitation windows (as further explained in SM Section S2.6), in analogy with strong-field ionization \cite{pabst16a}.

We therefore from hereon focus on the simplest model that reproduces the experimental observations -- the initial population of the B and C states only. The calculated electronic coherences are shown in Fig.~4B. The BC coherence appears at $\Delta t=0$, whereas the AC coherence progressively builds up, reaching a maximum around 7~fs. These results show that the electronic coherence between the B and C states is efficiently transferred to a coherence between the A and C states, mediated by the non-adiabatic population transfer from B to A.

\begin{figure}
    \centering
    \includegraphics[width=0.8\textwidth]{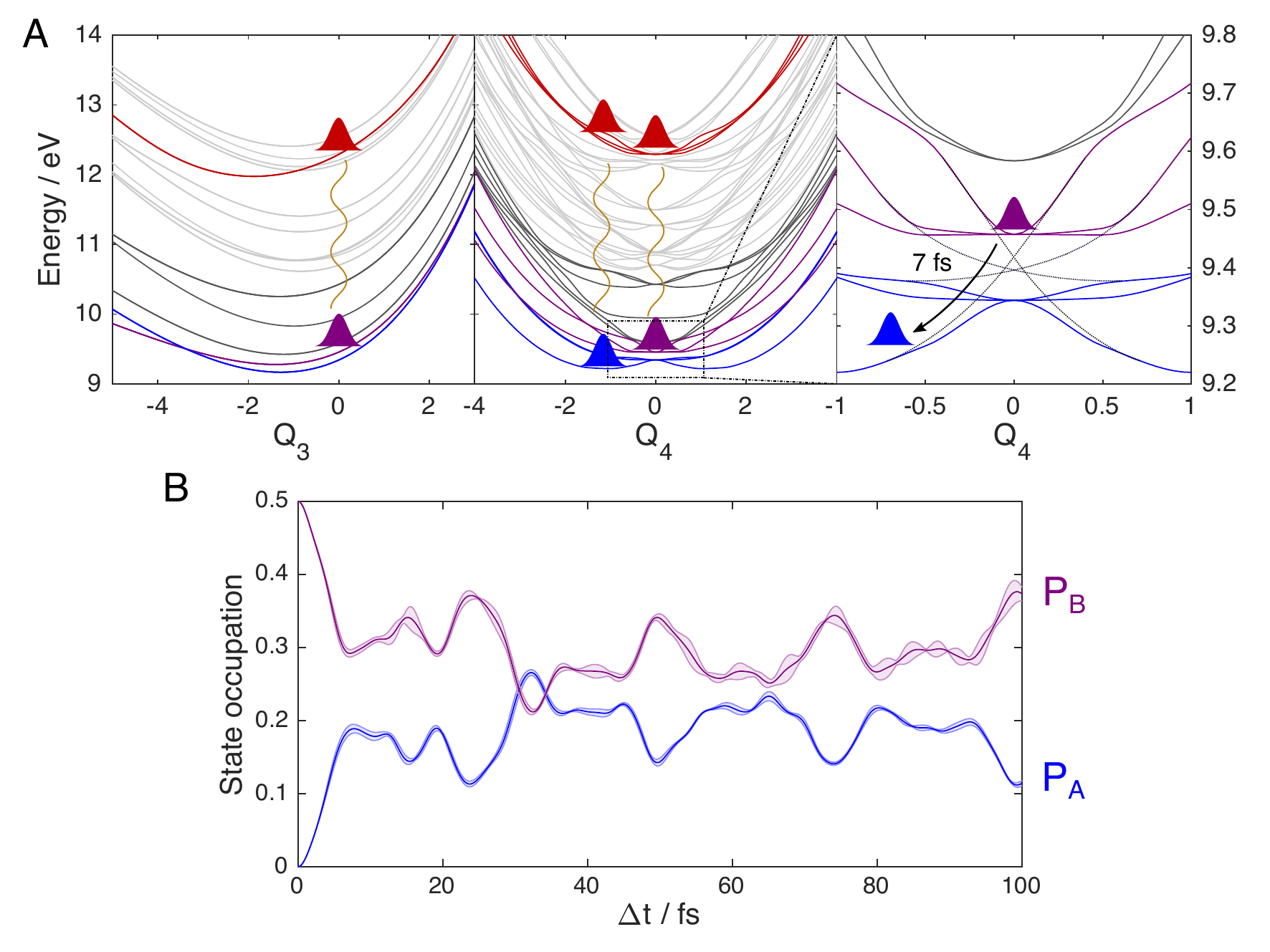}
    \caption{
    \textbf{Non-adiabatic transfer of electronic coherence} A) Adiabatic PES as a function of the symmetric ($\nu_3$) and anti-symmetric ($\nu_4$) stretching modes (left and middle). The curves have been colour-coded to reflect the electronic character of the corresponding states (see Fig.~2B for assignment). Dark-grey curves represent other states included in the MCTDH model while light-grey curves represent states excluded from the model. Magnified portion of the central panel with overlaid schematic diabatization (dotted lines) reflecting the conical intersection between the A and B states responsible for the population transfer (right). B) Diabatic MCTDH populations of the A and B states as a function of time after initialization.
    }
\end{figure}

\subsection{Attosecond Charge Migration and its Revival} \label{AttoCM}
\begin{figure}
    \centering
    \includegraphics[width=1.0\textwidth]{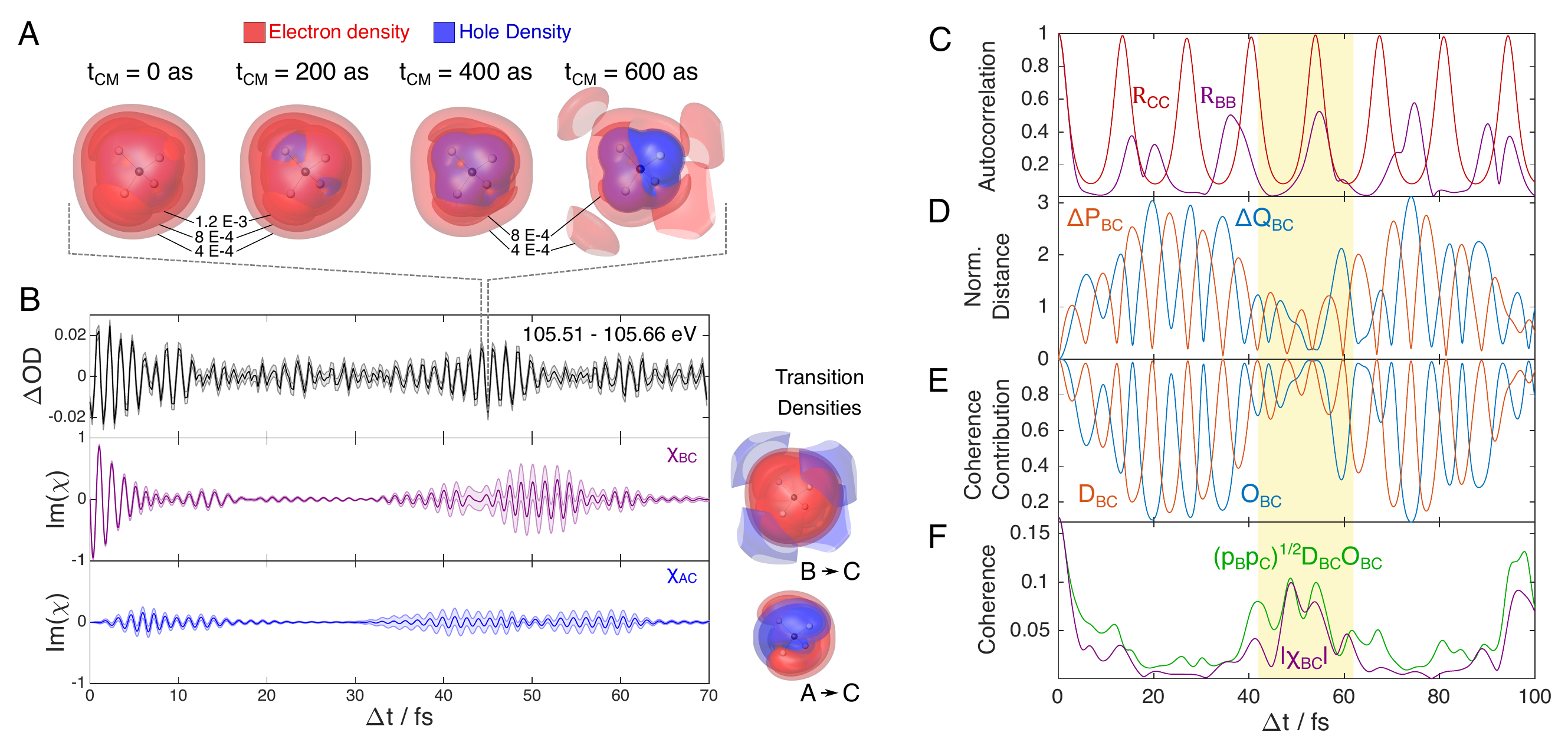}
    \caption{
    \textbf{Attosecond electron wavepacket, decoherence and revival.} A)  Electron-density difference between excited and unexcited molecules ($\rho_{\rm ES}(t)-\rho_{\rm GS}$) at selected delays covering half a quantum-beat period (as indicated by the dotted lines), isosurface values are labeled, B) Measured Fourier-filtered (0.5-1.5~PHz) $\Delta$OD (black, left scale) and MCTDH-calculated electronic coherences (purple/blue, right scale), the transition denisities related to each coherence are shown on the right-hand sied, C-F) Analysis of the decoherence and revival showing (C) the MCTDH auto-correlation functions of the nuclear wavepackets in the C-Rydberg (red) and B-valence (violet) states, (D) the difference between the position and momentum expectation values of the wavepackets in the B and C states, (E) their contributions to the mutual electronic coherence, and (F) the electronic coherence obtained from the semi-classical analysis (green) and from the full MCTDH calculation (violet). The revival has been highlighted in yellow for clarity.
    }
\end{figure}
We now turn to the detailed analysis of the attosecond charge migration and its evolution on the femtosecond time scale. Figure~4B shows the $\Delta$OD signal averaged over the 105.51-105.66~eV energy window. The signal displays the 1.31-1.39~fs period quantum beats that have been identified in Fig.~2. The initial decoherence of the quantum beat and the revival between 40 and 55~fs are also evident. A direct comparison between the amplitudes of the experimental signal oscillation and the calculated electronic BC and AC coherences shows excellent agreement up to a small offset in the timing of the maximal revival. %The calculations reproduce the dominance of the BC coherence experimentally observed at 0.720~PHz over the AC coherence detected at 0.765~PHz visible in the FFT.

The right-hand side of Fig.~4 elucidates the origin of this decay and revival of the BC coherence. Whereas it is clear that nuclear dynamics are responsible for the modulation of the coherence, the time scale of 50~fs is far longer than any vibrational periods included in the MCTDH model. As can be seen in Fig.~4C, the nuclear wavepackets in the different states complete multiple periods of vibration before the peak of the coherence revival. Whereas the dynamics on the C state are adiabatic and periodic, the wavepacket prepared on the B state exhibits non-adiabatic population transfer and therefore an autocorrelation function that does not return to its initial value. 

For coherence between two electronic states $i$ and $j$ to exist, they must both hold significant populations ($p_i,p_j$) and their vibrational wavepackets must have a non-zero overlap not only in coordinate space ($O_{ij} = e^{-\Delta Q_{ij}^2}$, where $\Delta Q_{ij}$ is the spatial separation), but also in momentum space ($D_{ij}= e^{- \Delta P_{ij}^2}$, where $\Delta P_{ij}$ is the momentum separation). The amplitude of the total electronic coherence can indeed be expressed as ($\sqrt{p_ip_j}O_{ij}D_{ij}$). Since these quantities are not directly accessible in the calculations, they were obtained within a Gaussian approximation to the vibrational wavepackets~\cite{golubev2020tga}; the corresponding quantities are shown in panels D and E (for details see SM, Section~S2.4). The product of the three terms ($\sqrt{p_{\rm B} p_{\rm C}} D_{\rm BC}O_{\rm BC}$) evolves similarly to the magnitude of the BC coherence, $\left|\chi_{\rm BC}\right|$, from the MCTDH calculation but tends to slightly overestimate the latter (panel F). Furthermore, Movie S2 shows that the source of this effect is mainly the $nu_3$ mode, as, unlike the $nu_4$ mode, its minimum is far from the Frank-Condon region, resulting in large-amplitude motion and significant separation of the respective vibrational wavepackets. Our calculations thus show that, due to the relative curvature of the $nu_3$ PES of the B and C states, it takes approximately 50~fs for the nuclear wavepackets to simultaneously regain coordinate- and momentum-space overlap, producing the revival. 

Having identified the experimentally observed coherences, we can combine the MCTDH and EOM-CCSD calculations to reconstruct the electronic density as a function of time. The spatial shape of the charge migration is defined by the transition densities ($\bra{\psi_i}\hat{D}\ket{\psi_j}$, where $\hat{D}$ is the EOM-CC density operator). The two most relevant ones are shown on the right-hand side of Fig.~4A. While the BC coherence causes charge to travel from the vicinity of the atoms to a diffuse cloud around the molecule, the AC coherence causes it to migrate between pairs of hydrogen atoms.

The time-dependent difference electron density with respect to the SiH$_4$ ground state resulting from applying the MCTDH coherences to these transition densities is shown in Fig.~4A. $t_{CM}=0$ has been chosen to correspond to times when the electron density is most strongly contracted around the molecule, resulting in an increase of core-Rydberg hole overlap and therefore the experimentally observed OD at 105.5 eV. The complete temporal evolution of charge migration is shown in a supplementary movie (Movie S1). The differently shaped transition densities can give rise to interesting phenomena - depending on the relative amplitude of the coherences ($\left|\chi_{ij}\right|$), the charge migration can transition from being radial to angular in nature, particularly noticeable at 40 and 60~fs in Movie S1. 

%The B-C transition density shows that charge can periodically migrate from a volume with a radius of 3.5~\AA, concentrated in the immediate vicinity of the atoms to a diffuse cloud with radius of over 4~\AA, while the A-C transition density, which is mostly confined to a volume with a radius $<$3.3~\AA, can represent electrons migrating between pairs of hydrogen atoms.
%The differently shaped transition densities can give rise to interesting phenomena - depending on the relative amplitude of the coherences ($\left|\chi_{ij}\right|$), the charge migration can transition from being radial to angular in nature, particularly noticeable at 40 and 60~fs. 
%These transitions are only evident from the simulations as the experimental results contain no direct spatial information, however, varying the relative polarisation of the pump and probe fields in such experiments could give evidence for such phenomena.

\subsection{Phase of Quantum Beats} \label{Phase}

Finally, we turn to the phase of the observed electronic quantum beats and analyze their information content. Figure~5A shows the transient spectrum and associated FFT phase of a coherent 3-level system. Although the phase of the quantum beat of the two lines is equal at the peak of the $\Delta$OD amplitude, the phase to the sides of these maxima are mirrored, producing a characteristic `U'-shaped phase profile. Such a signal is clearly distinct from few-femtosecond oscillations found in XUV-pump-IR-probe ATAS measurements, where the phase is initially flat and steepens towards larger negative delays (this is most evident when a Gabor filter is applied to the data, which is discussed in section S1.3 and visible in Fig.~S3).

Figure~5C shows the phase of the FFT of the BC coherence. In agreement with the 3-level system each quantum beat pair also 
displays mirrored phase profiles, identifying them as quantum beats. The much larger total variation of the phase in Fig.~5C compared to 5A is a result of the summation of the phase steps of spectrally adjacent core-valence transitions. The total phase variation therefore contains information, both on the number of final core-excited states and on the relative phases of the their transition-dipole moments, also known as the transition-dipole phase (TDP), a property of growing importance for attoscience \cite{Yuan2019}. The signature of the TDP on the quantum beats also carries an important physical meaning; out-of-phase quantum beats are the experimental manifestation of the migration of charge. When electron-hole density migrates, its overlap with the different Rydberg orbitals changes, periodically modulating the probability of different core-valence excitation.

To demonstrate this property, ATAS simulations \cite{golubev2020ATAS} have been performed based on the MCTDH results (Fig.~5B, see section S2.3 for details). The sensitivity of our experiment to the signs of the TDP is highlighted in Fig.~5B, which compares two simulations. The simulated transient spectra use the same absolute transition dipole values and transition energies for the A, B, and C to final core-excited state transitions, however, in the upper panel the signs of the transitions from the C state to the final states have been chosen to alternate as a function of energy (see Table S4), whereas equal signs have been used in the lower panel. Clearly, the sign alternation is required to reproduce the monotonic phase variation observed in the experiment. Moreover, the total phase variation is a signature of the number of final states. In our simulations, we have assumed 3 final states to be accessible from A, 7 from B, and all ten of them from C. This model calculation results in reasonable agreement with the experimental data, whereby the smaller calculated phase variation suggests that the number of final states is even larger in reality. This is not unexpected because of the doubly-excited (core+valence) character of the final states, but unfortunately beyond the capabilities of state-of-the-art quantum-chemical calculations. Nevertheless, this analysis demonstrates the sensitivity of ATAS to the TDP, an observable that is not accessible in conventional spectroscopy.

\begin{figure}
    \centering
    \includegraphics[width=\textwidth]{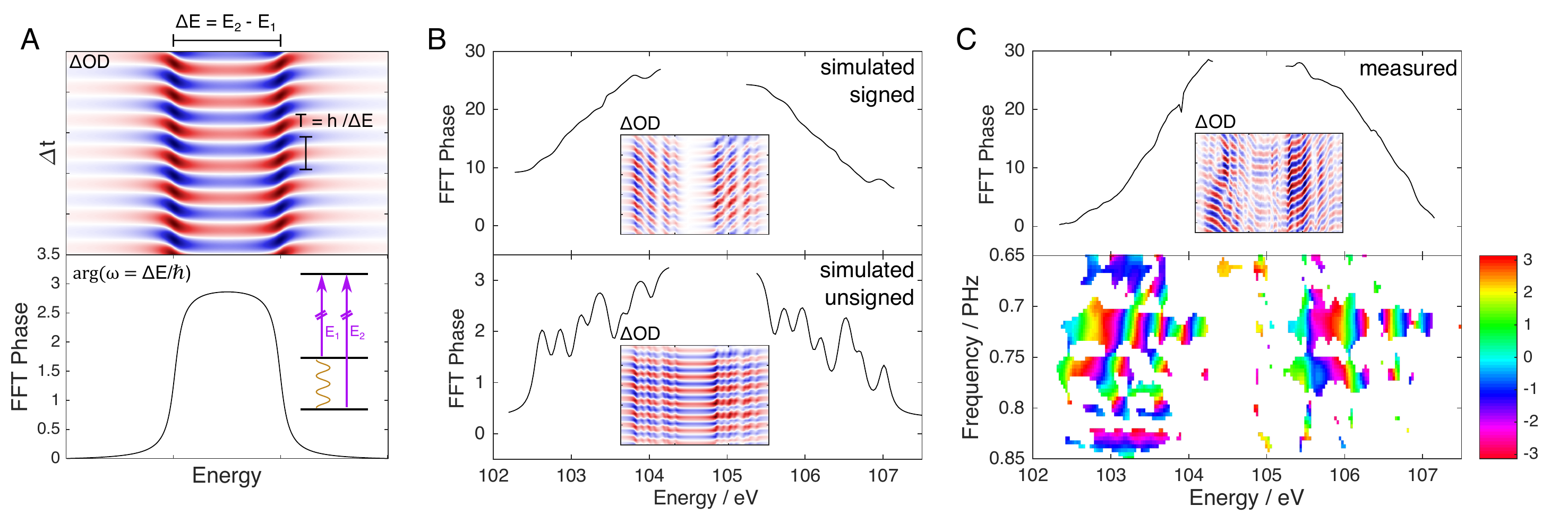}
    \caption{
    \textbf{Encoding of the sign of transition-dipole moments in molecular ATAS}  A) The spectral signature of a quantum beat in ATAS. B) Results of the ATAS simulations based on the MCTDH results. The top panel is obtained by selecting the relative signs of the transition dipoles to best match the experiment (Table S4), whereas the bottom figure is the result of choosing all transition dipoles positive. Simulated and measured ATAS results at the time of the revival are presented alongside the corresponding phase line-outs. C) Phase of the quantum beat extracted from the experimental transient spectrum. The bottom panel displays the phase of the FFT shown in Fig.~2A. Areas corresponding to negligible amplitudes have been left blank. An unwrapped line-out of this phase, taken at 0.72~PHz, is shown in the top panel with the experimental $\Delta$OD shown as inset.
    }
\end{figure}

\section{Conclusion}

This work has uncovered some remarkable opportunities in molecular attosecond spectroscopy. ATAS is capable of capturing specific electronic coherences in spite of the nearly featureless valence absorption spectrum (Fig.~2C). This is a direct consequence of the high temporal resolution on one hand, and the selectivity of the SXR-ATAS observables on the other. On the observed time scales, structural dynamics are limited to a subset of high-frequency stretching modes, which enables the preparation and survival of a complex electronic wavepacket that shows decoherence and revival. This information can not be obtained from the frequency-domain spectra, which are complicated by large-amplitude motions, driven by the Jahn-Teller effect that take place on longer time scales and involve the bending modes. On the investigated time scales, the quantum system can be thought of as evolving under the action of a simplified Hamiltonian that only contains the electronic and high-frequency vibrational modes, resulting in qualitatively simpler dynamics. The specificity of the SXR-ATAS observables arises from the projection of the dynamics under scrutiny onto common sets of Rydberg-like core-excited states. Since this probe step involves core excitations from and into orbitals with weak electron correlation, they should be analogous to those of the unexcited molecule; a similarity which is key to the straightforward assignment of the dynamics. The present method can thus be expected to favorably scale towards more complex molecules, since both the time-scale and the spectral-simplification arguments are general. 

Our results moreover show the possibility of preparing and observing charge migration in neutral molecules, while achieving a temporal resolution that lies significantly below the duration of the pump pulse. Specifically, we have observed the decoherence and revival of attosecond charge migration driven by structural and non-adiabatic dynamics. These results show that charge migration can not only survive coupled electron-nuclear dynamics, characteristic of excited-state dynamics in large molecules, but even be transferred to other electronic states. These observations also confirm the viability of the proposed control schemes over electronic degrees of freedom for manipulating chemical dynamics \cite{kling13a}. An immediate extension of our results is the control over attosecond charge migration \cite{golubev15a}, its application to steer charge transfer to a desired outcome and to thereby control chemical reactivity. All of these pillars of attosecond chemistry have now come within reach of experimental capabilities.

%%%%%%%%%%%%%%%%%%%%%%%%%%%%%%%%%%%%%%%%%%%%%%%%%%%%%%%%%%%%%%%%%%%%%%%%%%%%%%%%%%%%%%%%%%%%%%%%%
\section*{Acknowledgments}
We thank A. Schneider and M. Seiler for their technical support, D. Hammerland for laser support, D. Stefano for the coating of the Nb mirrors, J. Leitner and J. R. M\"o\ss{}inger for performing part of the test calculations, as well as V. Utrio Lanfaloni for the preparation of Fig.~1A. 
{\bf Funding} D.T.M. and H.J.W. gratefully acknowledge funding from the ERC Consolidator Grant (Project No. 772797-ATTOLIQ), and from the Swiss National Science Foundation through projects 200021\_172946 and the NCCR-MUST. V.D. and A.I.K. thank the DFG for the financial support provided through the QUTIF Priority Programme, and N.V.G. acknowledges the support by the Branco Weiss Fellowship---Society in Science, administered by the ETH Z\"urich.
{\bf Authors contributions} H.J.W. proposed the study. D.T.M. developed the experimental setup, performed the measurements and analyzed the data; V.D., N.V.G., and A.I.K. developed the theoretical models, and V.D. and N.V.G. carried out the calculations; H.J.W. supervised the experimental and A.I.K. the theoretical part of the project; H.J.W. and D.T.M. wrote the manuscript with input from all coauthors.
{\bf Competing Interests} None to declare.
{\bf Data and materials availability} All data needed to evaluate the conclusions in the paper are present in the paper or the supplementary materials.
%, as well as online at Zenodo \cite{}.

%%%%%%%%%%%%%%%%%%%%%%%%%%%%%%%%%%%%%%%%%%%%%%%%%%%%%%%%%%%%%%%%%%%%%%%%%%%%%%%%%%%%%%%%%%%%%%%%%
%Here you should list the contents of your Supplementary Materials -- below is an example.
%You should include a list of Supplementary figures, Tables, and any references that appear only in the SM.
%Note that the reference numbering continues from the main text to the SM.
% In the example below, Refs. 4-10 were cited only in the SM.
\section*{Supplementary materials}
Experimental methods\\
Theoretical modelling\\
Figs. S1 to S10\\
Tables S1 and S4\\
%References \textit{(1-...)}
\textbf{Caption for Movie S1:} Electron density difference between the excited and unexcited molecule ($\rho_{\rm ES}(t)-\rho_{\rm GS}$) as a function of the time delay since excitation, based on the results of the MCTDH and EOM-CCSD/aug-cc-pVTZ calculations. Nuclear motion is not displayed. The isosurfaces of the density difference have the same isovalues as in Fig.~3A. The periods of most intense attosecond charge migration are shown at a slower speed for clarity.\\
\textbf{Caption for Movie S2:} Projection of the vibrational wavepackets of all electronic states in the MCTDH model onto the $\nu_3$ symmetric and one $\nu_4$ asymmetric stretching modes. While the $\nu_4$ dynamics are clearly responsible for the diabatic population transfer, the wavepackets of this modes do not show significant motion and remain well overlapped arround the Frank-Condon region. Meanwhile, the $\nu_3$ dynamics are very periodic and show clear dephasing and then rephasing of the Rydberg and valence wavepackets around a delay of 50~fs. 

%%%%%%%%%%%%%%%%%%%%%%%%%%%%%%%%%%%%%%%%%%%%%%%%%%%%%%%%%%%%%%%%%%%%%%%%%%%%%%%%%%%%%%%%%%%%%%%%%
% Your references go at the end of the main text, and before the
% figures.  For this document we've used BibTeX, the .bib file
% scibib.bib, and the .bst file Science.bst.  The package scicite.sty
% was included to format the reference numbers according to *Science*
% style.
%BibTeX users: After compilation, comment out the following two lines and paste in
% the generated .bbl file.

%\newpage
%\begin{quote}
%{\bf REFERENCES AND NOTES}
%%\begin{enumerate}
%%
%%\item G. Gamow, {\it The Constitution of Atomic Nuclei and
%%Radioactivity\/} (Oxford Univ. Press, New York, 1931).
%%\label{dooley03b}
%%\item W. Heisenberg and W. Pauli, {\it Zeitschr.\ f.\ Physik} {\bf 56},
%%1 (1929).
%%
%%\end{enumerate}
%\end{quote}
%%
%%\bibliography{scibib}
%%\bibliographystyle{Science}

%\newpage
%\bibliography{attoH2On,RefSaijoscha}
\bibliography{attoH2On,attobib,biblio_SOM_Theory_Silane,extra}
\bibliographystyle{Science}

\end{document}

\baselineskip24pt

\maketitle

\noindent{\bf This PDF file includes:}\\
\noindent Experimental methods\\
Theoretical modelling \\
Figures S1 to S10 \\
Tables S1 to S4 \\
%References ...\\

\newpage
\tableofcontents

\renewcommand{\thefigure}{S\arabic{figure}}
\renewcommand{\thetable}{S\arabic{table}}
\renewcommand{\thesection}{S\arabic{section}}

\clearpage
\section{Experimental methods}

\subsection{Experimental setup}

An amplified carrier-envelope-phase (CEP) stabilized titanium:sapphire laser system (FEMTOPOWER V CEP) was used to deliver 25~fs, 1.5~mJ near-infrared (NIR) pulses centered at 790~nm with a full width at half maximum (FWHM) of 70~nm. These pulses were spectrally broadened in a 1~m-long, helium-filled hollow-core fibre. The gas pressure was adjusted such that the transmission efficiency of the fiber did not drop when helium gas was introduced, producing a spectrum spanning $>$350~nm with a central wavelength of 780~nm. Due to the third-order spectral phase imprinted onto the pulse by the broadening process \cite{jarque2018}, the pulse is compressible to 5.2~fs using broadband chirped mirrors and glass wedges (measured using a home-built, second-harmonic d-scan device). 

Once compressed, the $>$500~$\mu$J pulse was used to drive high-harmonic generation in a finite gas target filled with helium. A parabolic collimating mirror with a hole was used to separate the transmitted fundamental beam from its high harmonics, reflecting the diverging driving field while allowing the high-harmonic attosecond pulses to pass through. A 200~nm-thin zirconium foil was used to isolate the soft-X-ray pulse from the weak residual driving field before recombination. To avoid damaging the gold-coated toroidal refocusing mirror with the shorter-wavelength portion of the driving field, the filtering was performed directly in front of the toroid. In addition, two Nb$_2$O$_5$-coated mirrors were employed as partial beamsplitters to lower the driving field intensity, preventing the Zr filter from melting. A second parabolic mirror with a center hole placed after the toroidal allowed the attosecond high-harmonic pulse to pass through, collinearly recombining it with the recycled driving field while focusing the latter. The delay between the pump and the probe pulse is actively stabilised with a He-Ne interferometer and additionally monitored with white-light interferometry resulting in a delay jitter of $<$25~as. Additional details on the basic experimental setup are given in \cite{huppert15a}. 

The two pulses were focused into a 1~cm-long gas cell which was filled with the target gas. The spatial overlap between the pulses was optimized by maximizing the absorption signal from strong-field-ionized krypton. Time-zero was defined using the onset of the Kr$^+$ M$_{4,5}$-edge absorption signal. By recycling the driving pulse after high-harmonic generation, any additional nonlinear process that occurs inside the HHG target affects not only the generated harmonics but also the NIR pump pulse, which contributes to achieving an optimal temporal stability between the strong-field pump and the soft-X-ray (SXR) probe pulse. 

The transmitted spectrum was recorded using a CCD-based soft-X-ray spectrometer with sub-100~meV resolution at 100~eV, achieved by using a backlit in-vacuum CCD camera with a flat-field grating. The background signal from the residual driving field was minimized by using another 200-nm-thin Zr filter preceded by a pinhole which spatially removed a large portion of driving field, thus protecting the metallic filter from damage.

\subsection{Data acquisition and processing}

The active interferometric stabilization of our beamline allows the delay to be set to a specific value repeatedly. As such, each ATAS measurement performed on our beamline can include multiple scans of the same delay time resulting in a higher sensitivity. For the measurement presented in this work, two scans were preformed over a delay range of $-20$~fs to 70~fs with a step size of 250~as. For each scan and delay, exposures of 2 seconds were taken with the pump arm blocked and then unblocked. Before the measurement was started, a set of reference measurements were performed to establish the source of the different background signals and to allow for their subtraction. First, both the pump and probe arms were blocked and the dark counts of the CCD camera were captured. This exposure was used as a background for frames with the NIR arm blocked. Next, the probe arm was unblocked, providing a reference SXR spectrum to allow for the absolute absorption to be computed. Finally, since the pump pulse is sufficiently short and intense to produce harmonics in the target gas, only the pump arm was unblocked and the target filled with gas to provide a background for measurements with the pump arm unblocked.

To maximize the resolution of our spectrometer, as required to resolve the fine structure of the silane L$_{2,3}$ absorption spectrum, the curved nature of the spectral images produced by the flat field grating needs to be compensated for \cite{harada99a}. This is done by taking an average of the reference measurements over all delays and analyzing each row of the resulting reference image individually. For each row, the centre of mass of different absorption features is calculated. A cubic fit is performed to determine the position of the absorption features as a function of the vertical position on the CCD. This function is then used to shift the rows with respect to each other such that the absorption features line up with each other, allowing for a corrected, higher resolution spectral flux to be calculated. The horizontal pixel axis is converted into the photon energy axis by adjusting parameters in the grating equation until optimum overlap is achieved with the literature absorption spectrum of silane \cite{puettner97a}, resulting in Figure S1 (and Fig.~1C in the main text).

\begin{figure}
\includegraphics[width=0.85\textwidth]{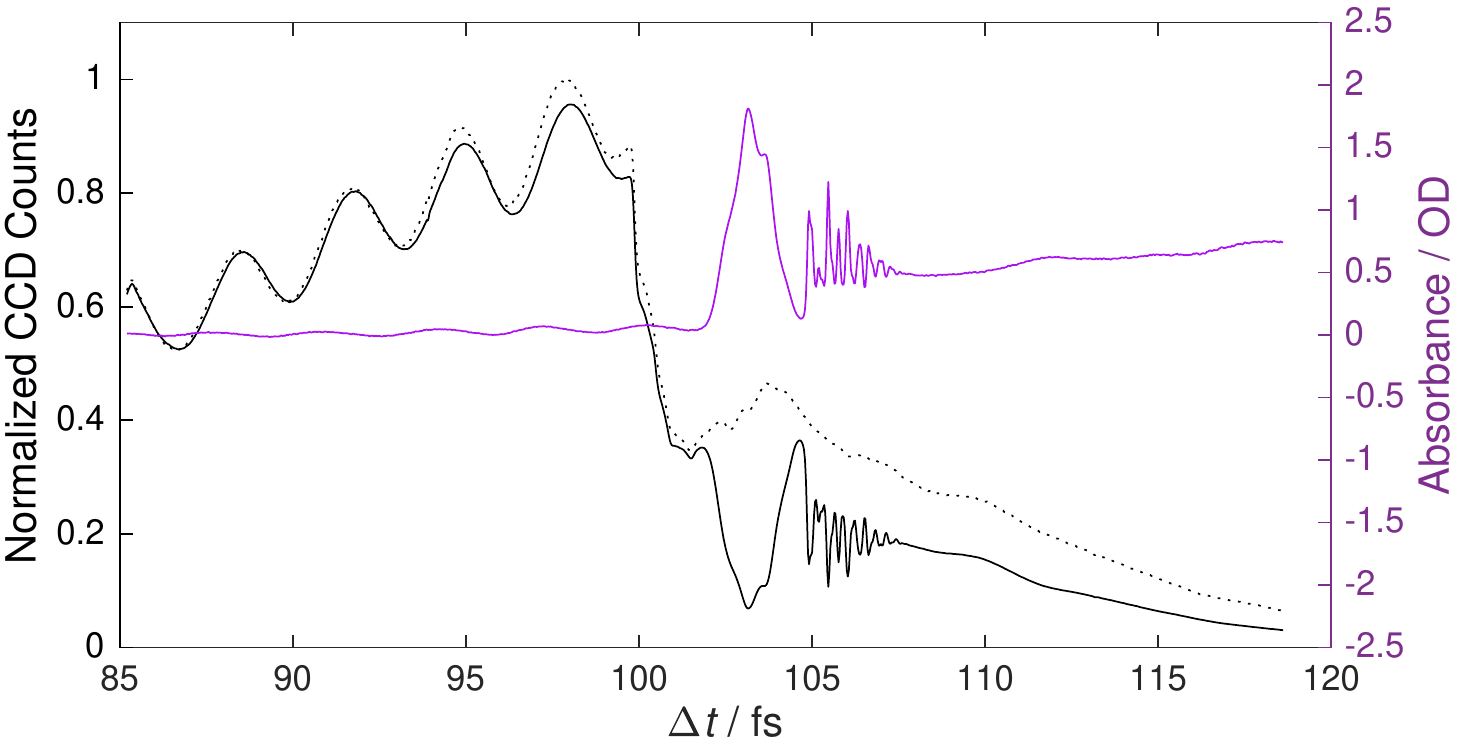}
\label{figs1}
\caption{\textbf{Attosecond soft-X-ray spectra and absorbance of silane.} The black dotted and solid lines are the measured probe-pulse spectra before and after the silane is introduced into the target gas cell, respectively. The absorbance of the unpumped silane gas can therefore be calculated and is shown as the purple line. 
%The wave-like structures in the spectrum are artefacts due to CEP fluctuations of the IR driving field. 
As the CCD detector is made of silicon, its quantum efficiency drops by a factor of 2 above $\sim$100~eV which is evident in the black lines. This effect is eliminated when calculating the absorbance and, therefore, does not affect the results shown in all other figures, including the main text.}
\end{figure}

Despite the CEP stabilization of the oscillator and amplifier, slight fluctuations in the pulse energy can still lead to fluctuations in the CEP of the few-cycle pulse, resulting in significant fluctuations of the probe pulse spectrum, especially in its cut-off region. These fluctuations happen on a sub-second time scale, therefore, subsequent exposures that measure the pump-induced change in absorption often have different spectra. The spectral signature of these CEP fluctuations is a shift in energy of the harmonic features in the spectrum, as well as a global change in the intensity of the harmonic flux. If left untreated when evaluating the absorption, the result, visible in Fig.~S2, is a random and significant fluctuation of the baseline around the harmonic cut-off, which transforms into a checkerboard-like pattern for photon energies that fall lower in energy (in our case for energies below 105 eV). Various solutions to this problem have been presented in the literature, making use of Fourier filtering \cite{stooss2019}, laser intensity referencing \cite{Volkov2019}, or taking advantage of spectral regions that are expected to not exhibit any changes in optical density (OD) \cite{timmers19}. 

\begin{figure}
\includegraphics[width=1.0\textwidth]{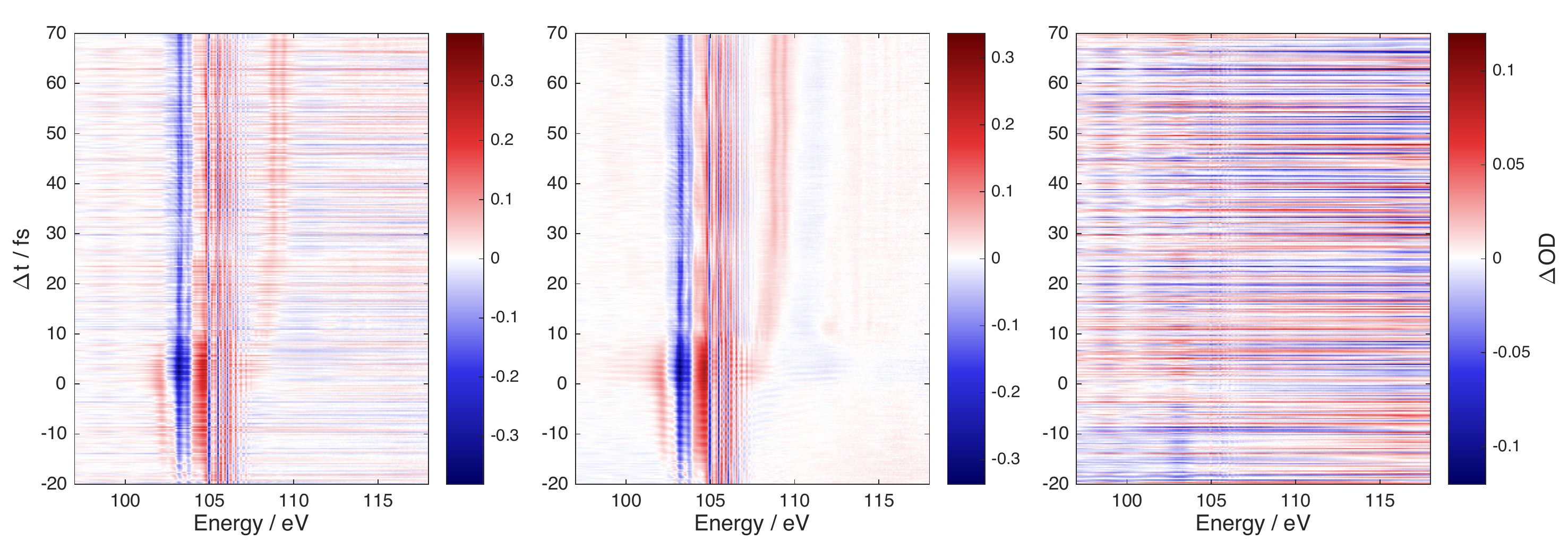}
\label{figs2}
\caption{\textbf{Overview of SVD filtering.} Unfiltered $\Delta$OD transient spectrum as obtained from the experiment (left), $\Delta$OD transient spectrum as obtained by removing four singular vectors among the leading ten singular vectors (center), and the residual of the filtering corresponding to the removed singular vectors (right).}
\end{figure}

In this work, we present an approach which takes advantage of the multiple scans that can be performed with our instrument. Based on the random nature of the fluctuations, we apply singular-value decomposition (SVD) to decompose our ATAS spectrum into orthogonal singular vectors which can be processed independently. The SVD can be seen as grouping correlated changes of the OD into the same singular vectors. Any changes to the measured OD that are caused by our pump pulse will have a well-determined relationship to the delay that shall be identical in subsequent measurements, while changes to the OD that are due to CEP fluctuations will instead have random fluctuations with respect to the delay. By performing the SVD on a matrix formed by concatenating subsequent scans rather than averaging over them, the singular delay vectors of different scans can be compared and their covariance can be evaluated. By setting a cutoff for the minimum allowed covariance, only singular vectors that have significant dependence on the delay can be selected and through simple matrix multiplication, be recombined to form an SVD-filtered ATAS spectrum which can now be averaged over the different scans. Noise has the effect of mixing the different singular vectors with each other, especially for those that have a small singular value, making it difficult to classify these with a simple covariance analysis. As a result, we have only filtered the first ten singular vectors, of which four were removed, and kept all singular vectors up to the 200th to avoid removing singular vectors with a small weighting in the ATAS spectrum but a significant effect. The result and the residual are shown in Fig.~S2. Once the singular vectors are obtained, they form an orthogonal space and therefore can be applied to any other spectrum. In this work, we have also performed the same processing to our static spectrum, shown in the main text (Fig.~1C), despite only having a single exposure of this spectrum. This technique has the advantage that no assumption is made about the spectral shape of the fluctuations nor the spectral regions that are expected to exhibit a signal, allowing the method to be applied to any system. In addition, all operations are linear which allows for rapid processing and transparent interpretation. However, one must be careful not to over-interpret the physical meaning of the individual singular vectors. 

\subsection{Time-Frequency analysis}

As is the case for many pump-probe techniques, ATAS experiments can invert the role of the pump and probe pulses in order to investigate either strong-field initiated dynamics, the focus of this paper, or instead, the interactions of core-excited states with a VIS/IR field. It is not possible to only induce one of these phenomena and, as a result, ATAS measurements often exhibit both. Due to the finite duration of the pulses, signals arising in the region of temporal overlap can originate from either phenomenon and require detailed analysis to classify and interpret them. As both phenomena can produce rapidly oscillating signals with few-femtosecond periods, a detailed time-frequency analysis is best able to separate the contributions. 

\begin{figure}
\includegraphics[width=1.0\textwidth]{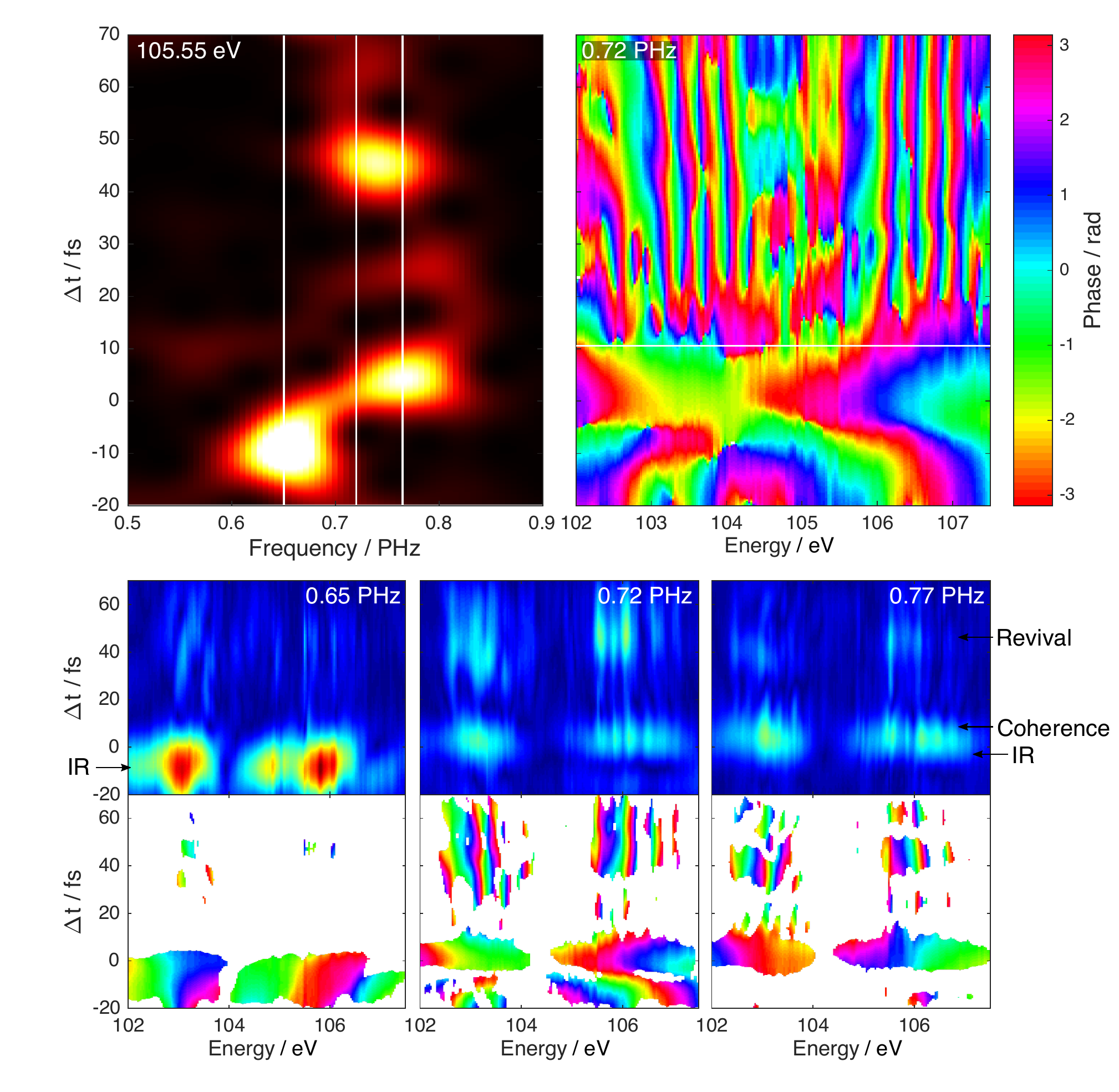}
%\label{figs3}
\caption{\label{Gabor}\textbf{Time-frequency analysis of the attosecond transient-absorption spectra.} The top-left panel shows the pseudo Wigner-Ville distribution of the $\Delta$OD at 105.55~eV. The vertical white lines indicate the three frequencies at which Gabor filters are applied to the entire transient-absorption data set. The results of these Gabor filters are shown in the three panels below. The amplitudes are shown above the phases, sharing a common color scale. Amplitude thresholding has been applied to the phases to ease the identification of the phase of the relevant signals. For completeness, the top right panel shows the unthresholded phase of the 0.72~PHz Gabor filter, exhibiting a pronounced change in structure for delays above 10~fs (highlighted by a white horizontal line). 
}
\end{figure}

The top-left panel of Figure~\ref{Gabor} shows a pseudo Wigner-Ville distribution of the $\Delta$OD signal at 105.55~eV -- the region of the spectrum exhibiting the largest-amplitude high-frequency oscillations. The strongest signal is found at negative time delays with a central frequency of 0.65~PHz. We can investigate the spectral distribution of this signal by applying a Gabor filter at 0.65~PHz on the transient spectrum (lower left panel of Fig.~\ref{Gabor}). This frequency is present over a broad spectral range at negative time delays, without exhibiting any distinct spectral peaks or large phase variations. These characteristics allow the assignment of this signal to the interaction of core-excited silane with the strong VIS/IR pulse \cite{chen13}. The coupling of light-induced sates (LIS) and/or dark states, with the bright states excited by the SXR pulse, results in a strong modulation of the final population of the core-excited state at twice the laser-field frequency. Although is 0.65~PHz is the second harmonic of a frequency that is red-shifted compared to the measured central frequency of our pump spectrum the significant third-order phase, introduced by the hollow-core fibre compressor and measured with a home-built second harmonic dispersion-scan setup (see Fig.~\ref{fig:FROG}E and F), results in a pre-pulse shoulder which exhibits a central frequency of 0.33~PHz, matching the frequency of the signal. 

From further examination of Fig.~\ref{Gabor}, we find, at positive time delays, oscillations with frequencies $>$0.7~PHz. Part of these signals originate from the same core-excited state -- IR coupling as discussed before, producing spectrally broad peaks with slowly varying phase. These signals dominate the Gabor phase up to a delay of 10~fs, however, the Gabor amplitude starts to exhibit narrow peaks already at these early delays. These narrow peaks are assigned to the excitation of a coherent superposition of electronic states into a common core-excited final state. As the pump and probe pulses lose temporal overlap and the initial decoherence of the electronic superposition sets it, both spectrally broad and narrow signals disappear. At this delay, the Gabor phase undergoes a significant change in structure. Due to the loss of pulse overlap and the IR now preceding the soft-X-ray pulse, no coupling of the core-excited states with the VIS/IR pulse is possible, resulting in the complete disappearance of the spectrally-broad signatures. This, in turn, makes the quantum beats of the electronic coherence the sole source of high-frequency oscillations in the transient spectrum, allowing the phase profile of the quantum beats to be observed. The phase exhibits two regions of monotonic (but step-like) increase/decrease, matching well the results of our ATAS simulations (see Fig.~5 of the main text), and supporting the assignment of the 0.7-0.8~PHz oscillations between 35 and 60~fs to the revival of the electronic coherence. As a result, the time window chosen to best isolate and represent the charge migration with minimal contribution of the LIS coupling was 10 -- 70~fs and is used for Fig.~4 of the main text. 

\subsection{Characterization of the pump pulse}

Unlike optical parametric amplification, HHG is an inefficient process that requires very high electric field intensities. As a result, multiple strategies have been devised to maximize the experimental conversion efficiency by recycling the optical driving field including HHG in enhancement cavities and split-mirror attosecond interferometers \cite{Bernhardt2012,Kienberger2004}. In the case of our experiment, we have made use of a rather unique technique which combines the efficiency of the split-mirror recycling with the advantages that separate pump and probe beamlines offer, which are normally achieved by placing beamsplitters before the HHG target. By placing a parabolic mirror with a drilled hole after the HHG target, the pump is split from the probe, simultaneously recycling the VIS/NIR while also allowing the beams to follow different paths where it becomes easier to spectrally shape the pulses. 

\begin{figure}
    \centering
    \includegraphics[width = 0.85\textwidth]{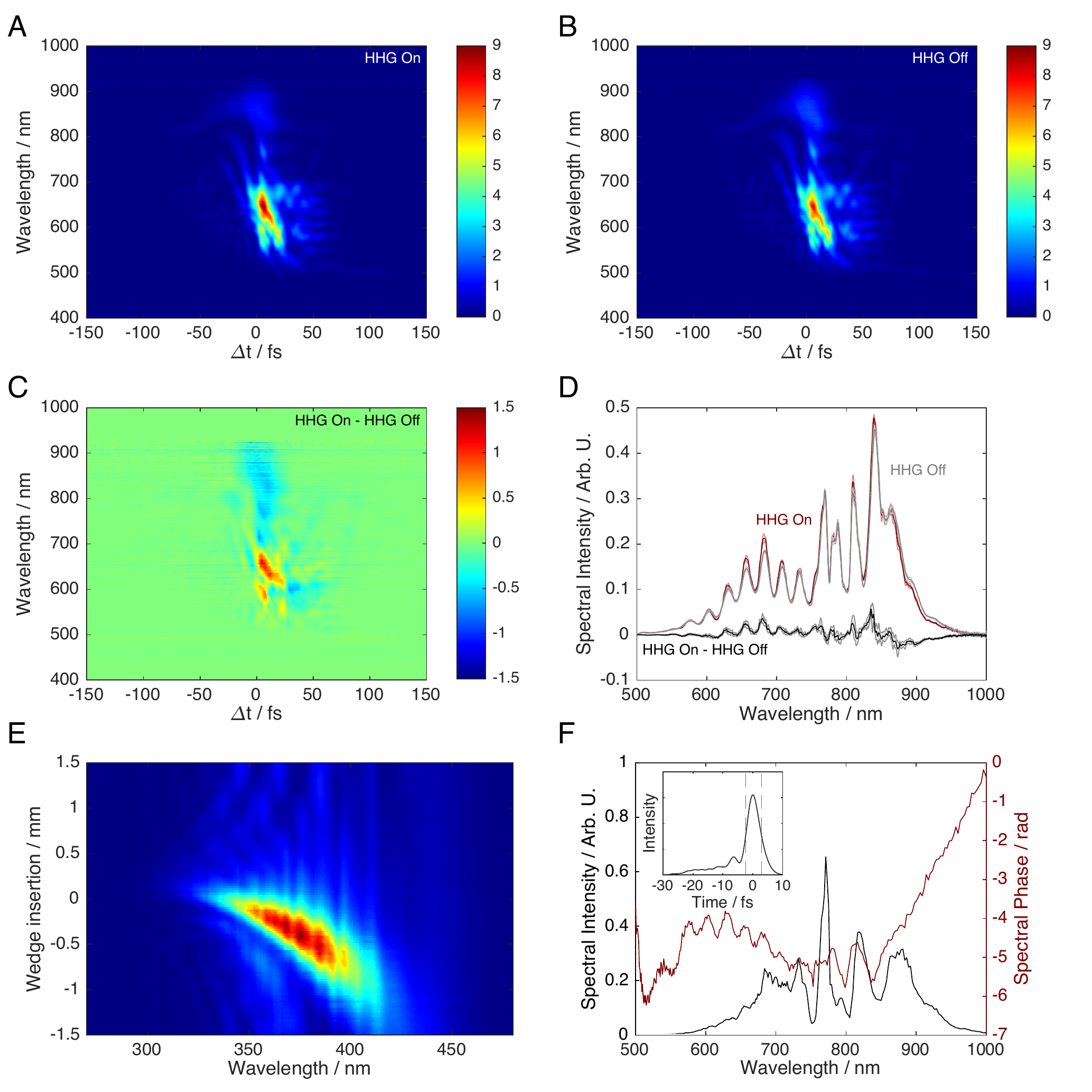}
    \caption{\textbf{Effect of HHG on the transmitted pump pulse}  A,B) TG-FROG measurements of the pump pulse with (A) or without (B) helium in the gas cell, taken after transmission through a 1~mm window (exit of the vacuum chamber). The difference of the two measurements is shown in C. D) effect of HHG on the spectrum of the transmitted IR few-cycle pulse. The the grey spectrum was measured without helium in the HHG gas cell, whereas the red was recorded with helium present. The effect, shown in black by taking a difference of the two spectra, is very small and is dominated by a $\sim$4~nm blueshift, consistent with panel C. E) result of second-harmonic-generation dispersion-scan (d-scan) measurement of the laser pulse performed in front of the experimental chamber. The residual third-order phase can be seen from the slanted SHG signal as well as in the reconstructed phase in F which also shows the reconstructed spectral intensity of the 5.2~fs pulse. The inset shows the temporal intensity profile of the pulse.}
    \label{fig:FROG}
\end{figure}

However, such a design also combines the disadvantages of the split mirror design (the creation of an annular pump-beam), and the separated beampaths (lower temporal stability). The issue of temporal delay between the pump and probe beams is resolved by employing a dual optical interferometer \cite{Huppert2015}. In this section we shall therefore discuss the effect of the HHG process and the drilled mirror on the pump beam. 

Unlike optical vortices, which exhibit a node at the centre of the beam's transverse mode in both the far and the near field, annular beams, which are created by applying a binary spatial mask to a regular Gaussian-like beam, do not exhibit a central node when focused. Therefore, the annular beam created by our drilled mirrors still exhibits a clean focus in our transient-absorption target, suffering only from a slight drop in maximum intensity and the generation of weak radial wings (using the convolution theorem we can think of this as arising due to the convolution of the unmasked focal spot with a sinc$^2$ function -- the Fourier transform of a top-hat function). A detailed analysis of these effects and their experimental application can be found in \cite{Gaumnitz2018}. The beam sizes, focal lengths, and mask sizes are almost identical to those used in our experiment.

The possible perturbation of the VIS/IR pulse in the HHG step is therefore the main source of concern. Although the process of HHG itself does not play an important part in reshaping the VIS/IR pulse, laser-plasma interactions of the driving field with free charges generated through strong-field ionization in the target are known and even exploited to increase the efficiency of HHG through phase matching. Nevertheless, it has also been shown that in cases of plasma defocusing intense enough to cause significant changes to the beam profile, HHG is not phase matched and no longer occurs \cite{Lai2011}. The use of 800~nm pulses and helium gas targets both reduce the rate of ionisation compared to longer wavelength drivers and other gases, therefore, such effects are most unlikely in our experiment. This is in agreement with the fact that no change in beam profile at the transient-absorption target is observed when performing HHG in helium.

While the spatial reshaping of the pulse by HHG is always weak, its spectral reshaping happens more readily and might also influence the results of our ATAS experiment. When HHG is performed in argon, the broadening of the spectrum is observed through self-steepening. On the other hand, helium, which is significantly harder to ionize and has an order-of-magnitude lower third-order susceptibility, exhibited no appreciable change in spectrum when the HHG gas was cycled on and off (see Fig.~\ref{fig:FROG}). The total change amounted to a blueshift of the spectrum by only $\sim$4~nm. Nonetheless, to further ensure that our pulses were not being modified by the HHG process, we performed transient-grating frequency-resolved optical-gating (TG-FROG) measurements on our pump pulses just before the transient-absorption target. Due to the fact that the pulse needs to be extracted from the experimental vacuum chamber in order to perform the TG-FROG measurement, it becomes positively chirped while passing through a window and therefore produces a rather complex FROG trace as shown in Fig.~\ref{fig:FROG}. When the HHG gas is cycled on and off, no significant changes to the FROG trace occur and only calculating the difference of the two FROG traces reveals any (Fig.~\ref{fig:FROG}A,B). In the difference plot (Fig.~\ref{fig:FROG}C) we see a slight intensity redistribution, however, the rich structure caused by the positive chirp remains completely unchanged, indicating that no significant changes to the spectral phase have occurred (Fig.~\ref{fig:FROG}D). For small redistributions of spectral intensity, one can expect negligible changes to the location of nodes in a TG-FROG trace, which is indeed the case in our results. We therefore conclude that the transmission of the few-cycle pump pulse the HHG medium has a negligible effect on its temporal structure. Finally, we also show a d-scan measurement and reconstruction of the few-cycle pulse performed in front of the vacuum chamber in Fig.~\ref{fig:FROG}E,F. The d-scan reconstruction in panel F agrees well with the measured spectrum in panel D. The temporal profile of the reconstructed pulse is shown in panel F with its 5.2~fs FWHM duration. The weak prepulse arises from the residual third-order chirp and has no influence on the dynamics reported in the main text, as discussed in section S1.3.

\section{Theoretical modelling}

\subsection{Excitation spectrum}

\begin{figure}
\begin{center}
\includegraphics[width=14cm]{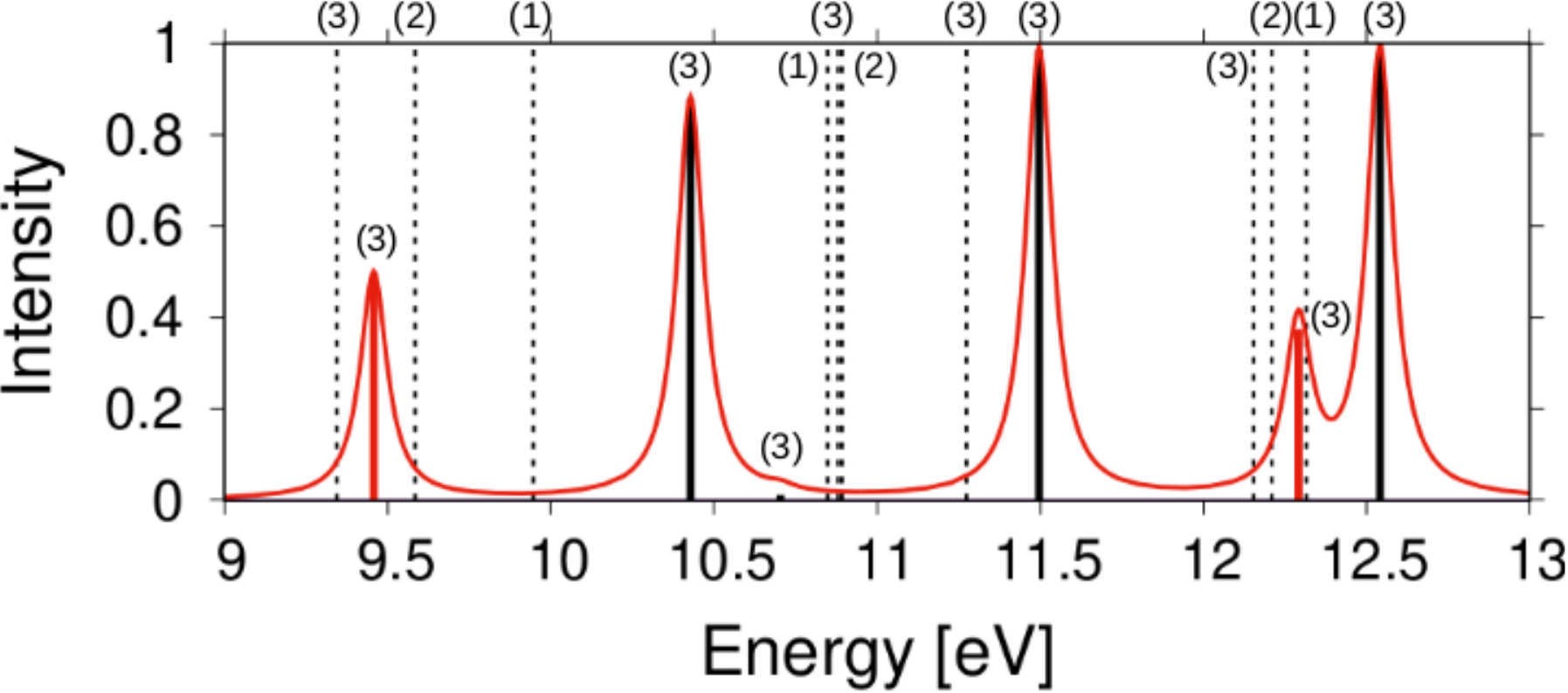}
\end{center}
\caption{\label{UVspec}\textbf{UV absorption spectrum of silane computed at EOM-CCSD/aug-cc-pVTZ level}. The intensity of each line reflects the square of the transition-dipole moment associated with the given state. The positions of the dark states obtained in this energy region are indicated with dashed lines and the degeneracy of each state is given in parentheses. The spectrum has been convoluted with a Lorentzian with FWHM of 50~meV to simulate the spectral broadening. The states that have been identified as responsible for the electronic coherence observed in the experiment are depicted in red.}
\end{figure}

The UV absorption spectrum of silane was computed at the EOM-CCSD/aug-cc-pVTZ level and is presented in Fig.~\ref{UVspec}, the list of the states can be found in Table~\ref{Tab_excitation}. Each line corresponds to an excited eigenstate of the system with an intensity determined by the respective transition-dipole moment from the electronic ground state. The spectrum is normalized to the strongest transition in this energy range. Due to the tetrahedral symmetry of silane, many of the states are doubly or triply degenerate. In addition, there are a number of dark states in this energy range, which cannot be populated directly from the ground state, since they have very small or zero (by symmetry) transition-dipole moments. In the experiment, however, the excitation is performed by a strong IR field, i.e., through a multiphoton absorption, and many of the single-photon-forbidden transitions will be nominally allowed. Nevertheless, we do not expect that the spectrum will change substantially, see also Sec.~\ref{Sec:SFE}. 

The spectrum was used to identify the states that could be responsible for the 0.72-0.765~PHz signal observed in the experiment. This signal has a period of 1.31-1.39~fs and thus corresponds to an energy splitting of 2.96-3.14~eV between the coherently populated states. Moreover, our analysis (see main text) shows that the observed beating is between a lower-lying valence state and an upper state with a Rydberg character from the $s$-series. A pair of states that satisfy these conditions is depicted in red in Fig.~\ref{UVspec}. The calculated energy gap between these two states is 2.83~eV, in good agreement with the experimental value. 

\begin{table}
\centering
\caption{\label{Tab_excitation}Electronic states present in the excitation spectrum shown in Fig.~\ref{UVspec}.}
\vspace{0.3cm}
\begin{tabular}{cccccc} 
 \hline \hline
 Numbering & Energy / eV & Degeneracy & Symmetry & Leading excitation 2t$_2\rightarrow$ \\ 
 \hline
 1 & 9.34   & 3 & T$_1$ & 3t$_2$ \\ 
 2 & 9.46   & 3 & T$_2$ & 4a$_1$ \\
 3 & 9.58   & 2 & E     & 3t$_2$ \\
 4 & 9.95   & 1 & A$_1$ & 3t$_2$ \\ 
 5 & 10.43  & 3 & T$_2$ & 3t$_2$ \\
 6 & 10.70  & 3 & T$_2$ & 4p (t$_2$) \\
 7 & 10.85  & 1 & A$_1$ & 4p (t$_2$) \\
 8 & 10.88  & 3 & T$_1$ & 4p (t$_2$) \\
 9 & 10.89  & 2 & E     & 4p (t$_2$) \\
 10 & 11.27 & 3 & T$_1$ & 3d (e) \\
 11 & 11.50 & 3 & T$_2$ & 3d (e) \\
 12 & 12.15 & 3 & T$_1$ & 3d (t$_2$) \\
 13 & 12.21 & 2 & E     & 3d (t$_2$) \\
 14 & 12.29 & 3 & T$_2$ & 5s (a$_1$) \\
 15 & 12.32 & 1 & A$_1$ & 3d (t$_2$) \\
 16 & 12.54 & 3 & T$_2$ & 3d (t$_2$) \\
 \hline
\end{tabular}
\end{table}

\subsection{Potential-energy surfaces and vibronic-coupling Hamiltonian}

In order to study the evolution of the system after the coherent population of this pair of states, we performed a quantum dynamics simulation taking into account the coupled motion of both electrons and nuclei. For this purpose, we used the well established and very powerful method of vibronic-coupling Hamiltonian \cite{koppel1984multimode}. In this approach, a Taylor expansion around a particular geometry point (usually the Franck-Condon point) is performed within a basis of quasi-diabatic states. The polynomial expansion is performed in terms of dimensionless normal coordinates, $Q$. The vibronic-coupling Hamiltonian can then be written as
\begin{equation}
\mathbf{H} = \pmb{\tau}_N + \pmb{\nu}_0 + \mathbf{W},
\end{equation}
where $\pmb{\tau}_N$ and $\pmb{\nu}_0$ denote the diagonal kinetic and the ground-state potential-energy matrices, respectively, while the matrix $\mathbf{W}$ contains the \textit{diabatic} states and the couplings between them.

Using a harmonic approximation for the vibrational modes, $\pmb{\tau}_N$ and $\pmb{\nu}_0$ can then be written as
\begin{eqnarray}
\pmb{\tau}_N = -\frac{1}{2}\sum_i\omega_i\frac{\partial^2}{\partial Q_i^2}\mathbf{1}, \\
\pmb{\nu}_0 = \frac{1}{2}\sum_i\omega_i Q_i^2\mathbf{1},
\end{eqnarray}
with $\omega_i$ being the frequency of mode $i$, and $\mathbf{1}$ denoting the unit matrix. In the present study, we used a quadratic model for the potential matrix $\mathbf{W}$, i.e., the Taylor expansion is truncated after the quadratic term, and, therefore, the matrix elements take the form
\begin{eqnarray}
W_{jj}&=&E_j+\sum_{i} \kappa_i^jQ_i+\frac{1}{2}\sum_{i}\gamma_i^jQ_i^2 \\
W_{jk}&=&\sum_{i}\lambda_i^{j,k}Q_i
\end{eqnarray}
In the above expressions, $E_j$ is the vertical excitation energy of state $j$, $\kappa_i^j$ and $\gamma_i^j$ are the linear and quadratic coupling parameters, respectively, of state $j$ for normal mode $i$, and $\lambda_i^{j,k}$ is the linear coupling parameter between states $j$ and $k$ by the normal mode $i$. The possible linear coupling parameters are determined through symmetry selection rules \cite{worth2004beyond}. The values of these parameters can be obtained through least-squares fitting to the \textit{adiabatic} potential energy surfaces (PES), obtained by solving the electronic eigenvalue problem at fixed nuclear geometry along the vibrational modes. With such a constructed vibronic-coupling Hamiltonian, we can propagate the nuclear wavepackets created by multiphoton excitation of the molecule by the strong IR pulse.

\begin{table}
\centering
\caption{\label{Tab1}Vibrational modes of silane.}
\vspace{0.3cm}
\begin{tabular}{ccc} 
 \hline\hline
 Energy / eV & Symmetry & Label \\ 
 \hline
 0.114 & t$_2$ & $\nu_1$ \\ 
 0.121 & e & $\nu_2$ \\
 0.276 & a$_1$ & $\nu_3$ \\
 0.277 & t$_2$ & $\nu_4$ \\ 
 \hline
\end{tabular}
\end{table}

\begin{figure}[ht!]
\centering
\includegraphics[width=0.7\textwidth]{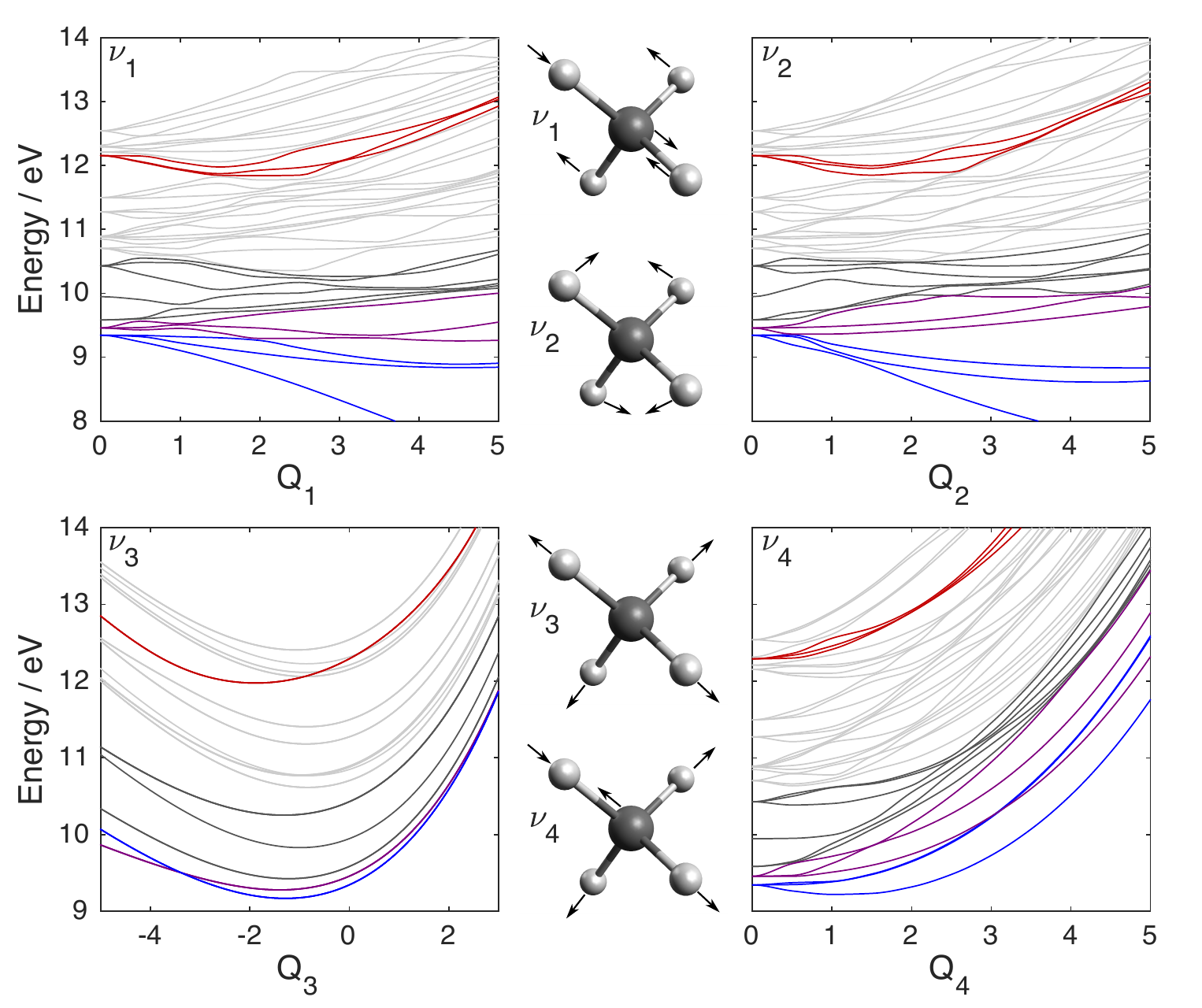}
\caption{\label{pes}\textbf{PES along the different vibrational modes of silane}. For the description of the modes, see Table~\ref{Tab1}. The color code is the same as in the main text, i.e., blue for the A-state, purple for B, and red for C. }
\end{figure}

The adiabatic PES along all the vibrational modes were computed at EOM-CCSD/aug-cc-pVTZ level of theory. The vibrational modes of silane are listed in Table~\ref{Tab1} and the PES along these modes are plotted in Fig.~\ref{pes}. 

The aim of our vibronic-coupling Hamiltonian model is to obtain a good description of the non-adiabatic dynamics initiated by the population of the pair of triply degenerate states identified previously and marked in red in Fig.~\ref{UVspec} (states B and C). Multiple tests, including different numbers of states and modes, were performed. We found that, due to its Rydberg character, the higher-lying triply degenerate state (C) is very weakly coupled to the other states and thus can be treated as isolated. This is in contrast to the lower-lying group of valence states. Due to the strong non-adiabatic coupling between these states, a good representation of the lower triply degenerate state (B) in the diabatic basis requires the inclusion of several other states, energetically lying both below and above it. Those states are depicted in blue (state A) and dark grey (others) in Fig.~\ref{pes}. As some of the states are degenerate, the final model includes 15 states, which are listed in Table~\ref{Tab2}. Due to the shapes of the PES along modes $\nu_1$ and $\nu_2$, it was not possible to obtain a satisfactory vibronic-coupling model accounting also for these vibrations and thus they were excluded from the vibronic-coupling Hamiltonian used to analyze the non-adiabatic dynamics of silane. As every degree of freedom can be a source of decoherence, the exclusion of some modes may in principle lead to an overestimation of the coherence time. As seen from Table~\ref{Tab1}, however, the frequencies of modes $\nu_1$ and $\nu_2$ are more than two times lower than $\nu_3$ and $\nu_4$ and thus are expected to have a much smaller influence on the dephasing of the initially created electronic coherence compared to modes $\nu_3$ and $\nu_4$. As we will see, this hypothesis is confirmed by the very good agreement of the theoretical results with the experimental observations. The vibronic-coupling Hamiltonian used to simulate the non-adiabatic dynamics triggered by the IR pulse, therefore, includes 15 states and 4 vibrational modes, the totally symmetric mode $\nu_3$ and the triply degenerate mode $\nu_4$.

\begin{table}
\centering
\caption{\label{Tab2}Electronic states present in the vibronic-coupling Hamiltonian and their calculated vertical excitation energies.}
\vspace{0.3cm}
\begin{tabular}{cccccc} 
 \hline \hline
 Numbering & Energy / eV & Degeneracy & Symmetry & Leading excitation 2t$_2\rightarrow$ & Label used\\ 
 \hline
 1 & 9.34   & 3 & T$_1$ & 3t$_2$ & A\\ 
 2 & 9.46   & 3 & T$_2$ & 4a$_1$ & B\\
 3 & 9.58   & 2 & E     & 3t$_2$ & \\
 4 & 9.95   & 1 & A$_1$ & 3t$_2$ & \\ 
 5 & 10.43  & 3 & T$_2$ & 3t$_2$ & \\ 
 14 & 12.29 & 3 & T$_2$ & 5s(a$_1$) & C\\ 
 \hline
\end{tabular}
\end{table}

\subsection{Non-adiabatic dynamics and construction of transient-absorption spectrum}

The vibronic-coupling Hamiltonian described above was used to propagate the nuclear wave packets on the coupled manifold of diabatic electronic states with the help of the Multi-Configuration Time-Dependent Hartree (MCTDH) method \cite{meyer1990multi}. MCTDH is a powerful grid-based method for numerical integration of the time-dependent Schr\"odinger equation, particularly suitable for treating multidimensional problems \cite{meyer2009multidimensional}. The Heidelberg MCTDH package \cite{mctdh} was used for the present calculations. 

Two types of calculations were performed. In the first type of calculations, the initial state was chosen to be a coherent superposition of the states denoted as B and C in Table~\ref{Tab2}. The results of these calculations are shown in Figs.~3 and 4 of the main text and used to analyze and interpret the experimental observations. To account for the possible population of other states during the multiphoton excitation step, a second type of calculations were performed, whereby all 15 states shown in Table~\ref{Tab2} were equally populated. The results of the latter calculations will be discussed below.

The coherence between each pair of electronic states can be extracted from the time-dependent overlap of the nuclear wave packets evolving on the corresponding PES (see, e.g., Ref.~\cite{despre2018charge})
\begin{equation}
\label{eq:coherence}
\chi_{jk}(t)=\int d \mathbf{Q}\, \chi^*_j(\mathbf{Q},t) \chi_k(\mathbf{Q},t),
\end{equation}  
where $\chi_j(\mathbf{Q},t)$ is the time-dependent nuclear wave packet, corresponding to state $j$, and $\mathbf{Q}$ denotes the nuclear degrees of freedom. The quantities $\chi_{jk}(t)$ are also important ingredients for computing the attosecond transient absorption spectrum, needed for having a direct comparison between theory and experiment. Using time-dependent perturbation theory and the Condon approximation, the ATAS can be obtained via the following expression for the transient absorption cross section (for a complete derivation, see the Supplemental Material of Ref.~\cite{golubev2020ATAS}): 
\begin{equation}
\label{eq:ATAS_cs}
\begin{split}
  \sigma (\omega,\tau) = \frac{4\pi\omega}{c} 
  & \text{Im} \sum_j \sum_k 
    \chi_{jk}(\tau)
    \sum_f \langle \Phi_j | \hat{\mu} | \Phi_f \rangle
           \langle \Phi_f | \hat{\mu} | \Phi_k \rangle \\ & \times \left(
      \frac{1}{\tilde{E}_f - E_j - \omega}
    + \frac{1}{\tilde{E}^*_f - E_k + \omega}
    \right),
\end{split}
\end{equation}
where $\omega$ is the photon energy, $\tau$ is the delay between pump and probe pulses, $c$ is the speed of light in vacuum, $\langle \Phi_j | \hat{\mu} | \Phi_f \rangle$ and $\langle \Phi_f | \hat{\mu} | \Phi_k \rangle$ denote transition-dipole matrix elements between initial $\{j,k\}$ and final $f$ electronic states with corresponding energies $E_{\{j,k,f\}}$ computed at the equilibrium nuclear geometry $\mathbf{Q}=0$. When $j=k$, the nuclear wave-packet overlaps $\chi_{jk}(\tau)$ give the populations of the initial electronic states. The energies of the final states are chosen complex, $\tilde{E}_f = E_f - i\frac{\Gamma}{2}$ to account for the spectral broadening. 

\begin{figure}[h!]
\includegraphics[width=1.0\textwidth]{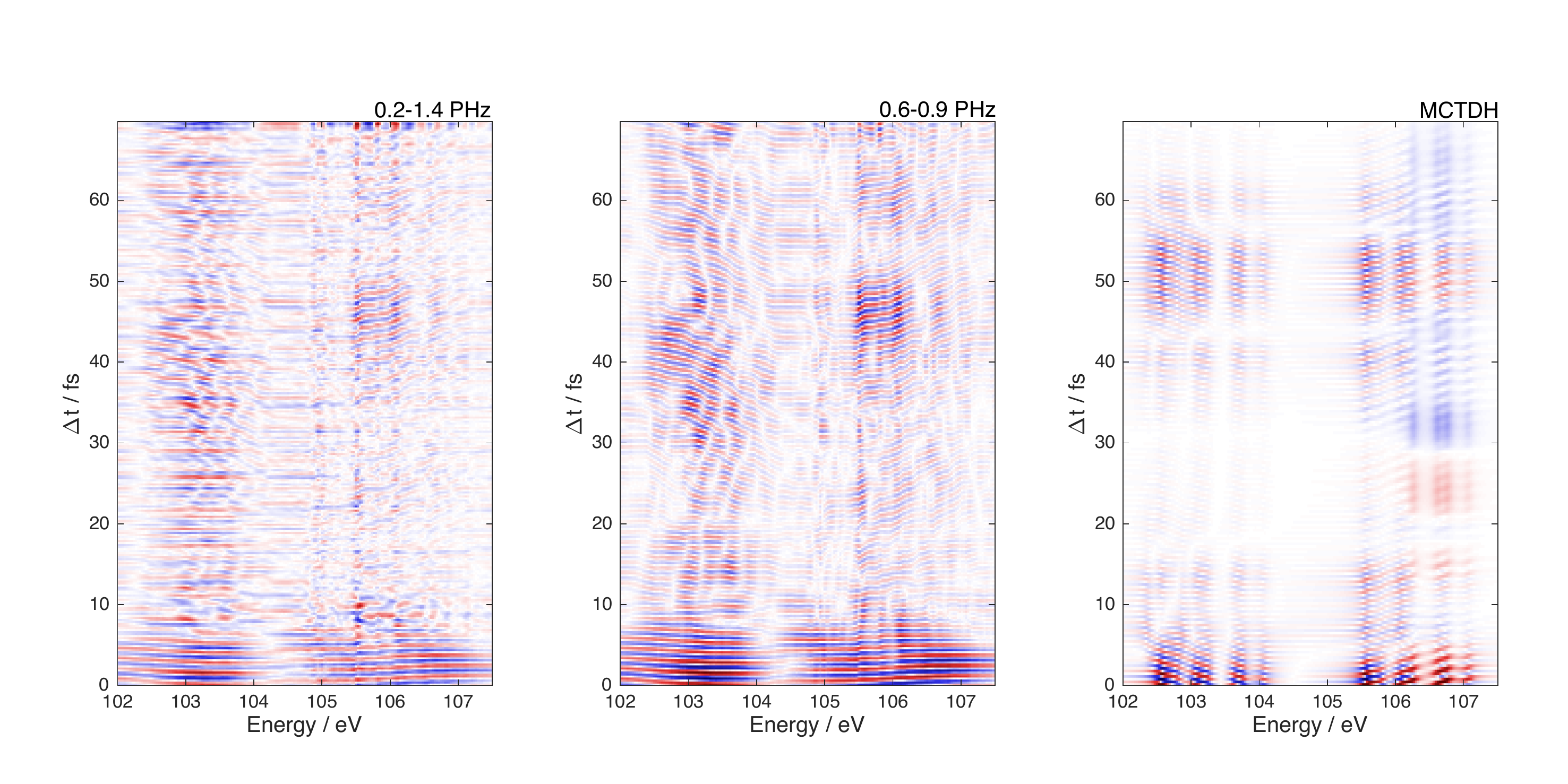}
%\label{figs3}
\caption{\label{comparison}\textbf{Comparison between Fourier-filtered $\Delta$OD and MCTDH calculations.} The experimental transient spectrum has been Fourier-filtered to suppress signals resulting from purely vibrational dynamics and is shown in the first two panels. The signal frequencies left after the application of the Fourier filter are displayed in the top-right; the second panel has been passed through a narrower acceptance range filter and thus exhibits less noise. The right-most panel shows a simulated transient spectrum based on the results of the MCTDH calculations. The average of the calculated spectra at each photon energy has been subtracted from the simulation to allow for an easier comparison between the three figures.}
\end{figure}

In order to obtain the ATAS, the transitions between the initial and final states need to be determined. The initial states consists of the states present in the vibronic-coupling Hamiltonian that get a significant population during the MCTDH propagation. As the ground-state equilibrium geometry of the neutral system is assumed for the simulation of the ATAS, the degeneracy is not lifted and a single initial state can be associated to each group of degenerate states. Therefore, the three initial states included in the simulation of ATAS are the ones located at 9.34~eV (A), 9.46~eV (B), and 12.29~eV (C), respectively. As core excitations from Si(2\textit{p}) are used in the experiment to probe the non-adiabatic dynamics in those states, the set of final states is formed by core excitations from Si(2\textit{p}). %Assuming that the valence excitation has a negligible impact on the core transition, and focusing on the specific transition at 105.5~eV, the final states are obtained by shifting each initial one by 105.5~eV. Therefore, the ensemble of final states consists of three states located at 114.84 eV, 114,96 eV and 117.79 eV.

As we were not able to compute the transition-dipole moments between the initial valence-excited states and the doubly-excited final states, their values were determined using the experimental results. We want to point out, however, that the particular values that were chosen will not have an impact on the probed dynamics in the valence-excited region, but will only change the respective intensities of the different lines and the energy they appear at in the transient spectrum. The obtained theoretical ATAS is plotted in the right panel of Fig.~\ref{comparison}. The experimental results are presented in the two left-hand plots of the figure for comparison. We see that the ATAS trace reproduces the main experimentally observed features well. The early-time coherence, manifesting itself as a quantum beating in the absorption of the initially populated states, is lost in about 10~fs and a revival is observed at around 50~fs. 

\begin{table}
\centering
\caption{\label{Tab3}Final electronic states $f_i$ and transition dipoles from the valence-excited states (A,B,C) used for computing ATAS. The negative signs in the right-hand column indicate which transition dipole elements were negated to reproduce the correct phase relationship between quantum beats. The units of the transition dipoles are arbitrary and only the relative signs and magnitudes matter.}
\vspace{0.3cm}
\begin{tabular}{ccccc} 
 \hline \hline
 Energy / eV & A $\to$ F$_i$ & B $\to$ F$_i$ & C $\to$ F$_i$ \\ 
 \hline
 105.35      & 1.0       & 0.0 &    0.7 \\
 105.60      & 0.7       & 0.0 & -0.5 \\
 105.85      & 1.0       & 0.0 &    0.7 \\
 \hline
 105.55      & 0.0       & 1.0 &    0.7 \\
 105.80      & 0.0       & 0.7 & -0.5 \\
 106.05      & 0.0       & 1.0 &    0.5 \\
 106.25      & 0.0       & 0.7 & -0.5 \\
 106.60      & 0.0       & 0.7 &    0.5 \\
 106.75      & 0.0       & 0.7 & -0.5 \\
 107.05      & 0.0       & 0.5 & -0.5 \\
 \hline
\end{tabular}
\end{table}

Our analysis shows that the second coherence is created after the wave packet propagating on the initially populated state at 9.46~eV (B) is transferred to the lower-lying dark state (A) through conical intersections (CI) located close to the Franck-Condon point. These results are particularly important, as conical intersections were expected to destroy the electronic coherence \cite{arnold2017electronic}. We see, however, that in this case, not only is the initial coherence not destroyed by the splitting of the wave packet at the CI, but a new one is created between the state populated through the CI and the upper Rydberg-like state (C). These results show that the effect of conical intersections on the coherence is case-dependent and can destroy \cite{arnold2017electronic}, conserve, and even create \cite{kowalewski15a} new coherences in the system. 

\begin{figure}[h!]
\includegraphics[width=1.0\textwidth]{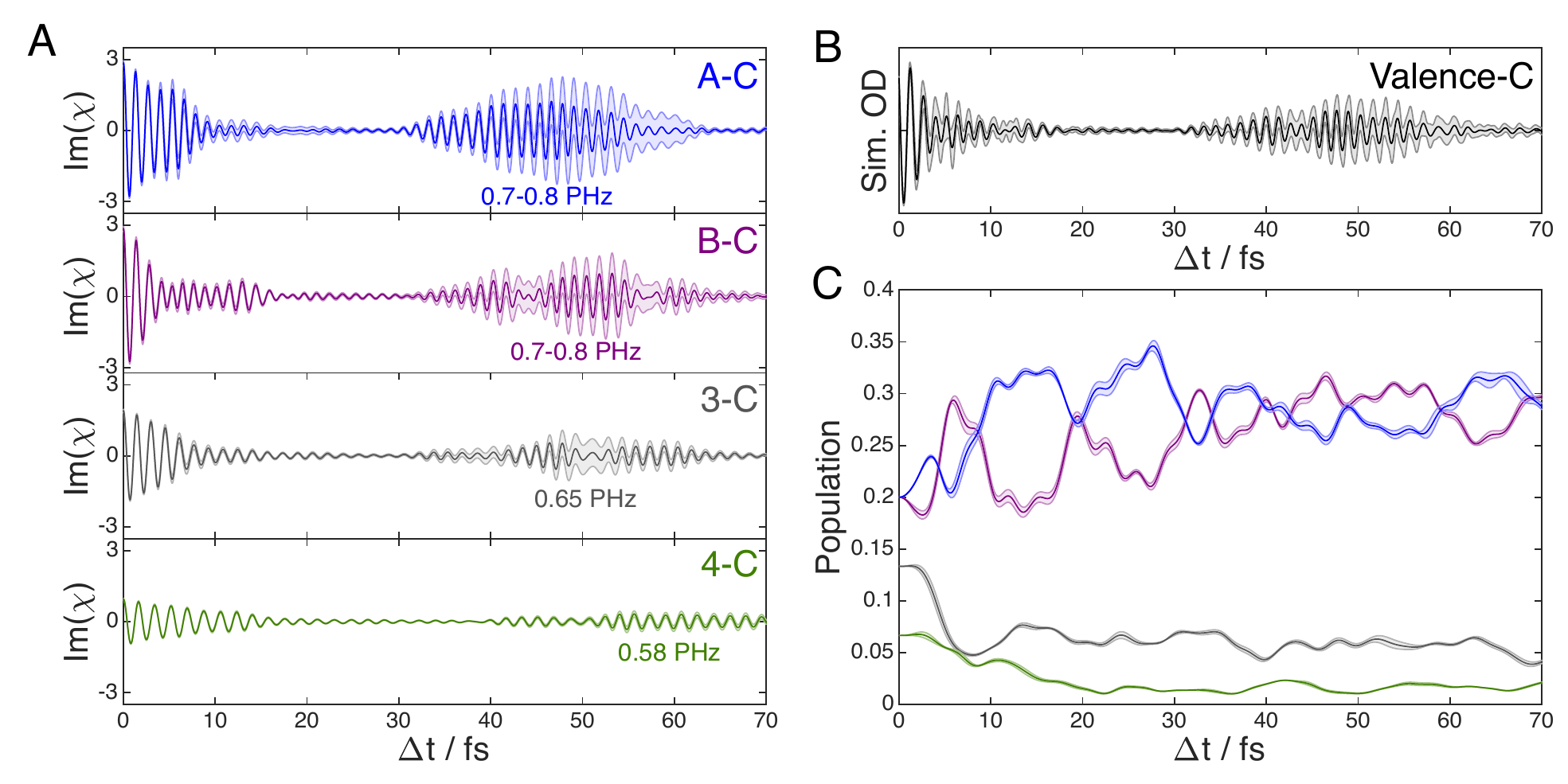}
\caption{\label{robustness}\textbf{Influence of other electronic states and their initial phases on the electronic dynamics} Here, we present the results of the first limiting case (mentioned in the main text), in which all states were initially populated. The robustness of the predicted coherence and population dynamics with respect to variations of the initial phase was investigated by averaging the results over 10 phase-randomised simulations. A: Coherence between states A (blue), B (purple), 3 (grey), or 4 (green) with state C. B: Simulated $\Delta$OD at 105.5 eV obtained by coherently adding all coherences between the valence states and the C state. C: Population dynamics of states A (blue), B (purple), 3 (grey), and 4 (green).}
\end{figure}

As the exact state of the system after the pump-pulse is unknown and, in addition, may vary from shot to shot and over the interaction volume, we found it important to verify that the MCTDH results were robust with respect to changes to the initial conditions. Robustness to the initial population of the states is established by performing MCTDH simulations in which all states (and degenerate sub-states) in the model are initially equally populated. As is the case for the simulations shown in the main text (Figs. 3 and 4), the robustness with respect to the initial phase of the wavepackets is also simultaneously investigated by performing ten separate simulations in which the initial phase is randomized. Randomizations that show an initial coherence drop of $<$20\% are selected for propagation. This criterion restricted the total range of the phase randomisation to 600~mrad. The result of the ten simulations are shown in Fig~\ref{robustness} where they are coherently averaged.

Figure~\ref{robustness}C shows that the population dynamics of the system quickly approach those of the calculation in which only the B and C states were initially populated ("BC calculation"). This is a consequence of the strong non-adiabatic coupling, which induced rapid population transfer from states 3 and 4 to states A and B within the first 10~fs. After about 20~fs, both simulations show that the majority of the valence population is in the A and B states, and is nearly equally distributed between them. As in the BC calculation, the simulation with all states initially populated shows that the coherence is preserved after the non-adiabatic population transfer and a revival of the quantum beat takes place between 40 and 50~fs. In fact, the revival in Fig~\ref{robustness}B occurs earlier than in Fig.~4B, while still being dominated by the B-C and A-C coherences, giving an even better agreement with the experimental results. The lower amplitude and weaked and/or delayed revivals of the 3-C and 4-C coherences is clear in Fig~\ref{robustness}A. 

These results therefore show that the silane molecule is very efficient at producing A-C and B-C quantum beats, a property that derives from the high density of CIs that funnel the populations of the valence-excited states into the lowest-lying excited states, to the quasi-degeneracy of its stretching modes and, most importantly, to the clock-like nature of the dynamics in the C-state, which lacks significant coupling to the other Rydberg states. The robustness of the results, combined with the excellent agreement between theory and experiment, serve as an \textit{a posteriori} validation of the hypothesis that our vibronic-coupling Hamiltonian model, even with its reduced dimensionality, is able to describe well all important features of the non-adiabatic dynamics of the system initiated by the IR pulse. 

\subsection{Semi-classical analysis of electronic coherences} 
 
In order to understand the mechanism of dephasing and revival of the electronic coherence observed in our experiment and reproduced by the accurate MCTDH simulations, we performed a theoretical analysis using a simple semi-classical model reported previously in Ref.~\cite{golubev2020tga}. We approximate the quantum wave packet $\chi_{j}(\mathbf{Q},t)$ propagating on the corresponding electronic state $j$ by a single Gaussian function
\begin{equation}
\label{eq:GWP}
\chi_{j}(\mathbf{Q},t) \approx \chi^G_{j}(\mathbf{Q},t) = 
    c_j(t) N \exp\left\{
        -\frac{1}{2\hbar}[\mathbf{Q}-\mathbf{Q}_{j}(t)]^{2}
        +\frac{i}{\hbar}\mathbf{P}_{j}(t)^{T}\cdot\lbrack\mathbf{Q}-\mathbf{Q}_{j}(t)]
        +\frac{i}{\hbar}\gamma_{j}(t)
    \right\},
\end{equation}
where $\mathbf{Q}$ and $\mathbf{P}$ are mass- and frequency-scaled coordinates and momenta, $\gamma_{j}(t)$ is the time-dependent phase of the wave packet, $N$ is the normalization constant, and $c_j(t)$ represents the time-dependent population of the electronic state. To establish the correspondence between the quantum wave packet $\chi_{j}(\mathbf{Q},t)$ and its semi-classical representation $\chi^G_{j}(\mathbf{Q},t)$, we set the position $\mathbf{Q}_j$ and momentum $\mathbf{P}_j$ of the Gaussian by computing the expectation values of the corresponding operators for the quantum wave packet evolving on the electronic state $j$
\begin{eqnarray*}
    &&\mathbf{Q}_j = \int d \mathbf{Q}\, \chi^*_{j}(\mathbf{Q},t) \mathbf{Q} \chi_{j}(\mathbf{Q},t),\\
    &&\mathbf{P}_j = -i \hbar \int d \mathbf{Q}\, \chi^*_{j}(\mathbf{Q},t) \nabla \chi_{j}(\mathbf{Q},t).
\end{eqnarray*}
The simple Gaussian form of the wave packet, Eq.~(\ref{eq:GWP}), allows us to express the electronic coherence between states $j$ and $k$, Eq.~(\ref{eq:coherence}), analytically as
\begin{equation}
\label{eq:SC_coh}
    \chi_{jk}(t)=c_j(t) c_k(t) e^{-d(t)^{2}/4\hbar}e^{iS(t)/\hbar},
\end{equation}
where
\begin{equation}
\label{eq:ph_dist}
    d(t)=\sqrt{|\mathbf{Q}_{j}(t)-\mathbf{Q}_{k}(t)|^{2}+|\mathbf{P}_{j}(t)-\mathbf{P}_{k}(t)|^{2}}
\end{equation}
is the distance in coordinate and momentum space between the centers of the two Gaussian wave packets and $S(t)$ is the phase responsible for the electronic oscillations. Further, we will focus on the absolute value of the electronic coherence, i.e., on the quantity $|\chi_{jk}(t)|$, thus omitting the consideration of the phase $S(t)$.

It is seen from Eqs.~(\ref{eq:SC_coh}) and (\ref{eq:ph_dist}) that the electronic coherence between a pair of electronic states $j$ and $k$ can be decomposed in three different components~\cite{vacher2017}: (I) the product of populations of the corresponding electronic states, (II) the overlap of the two nuclear wave packets in coordinate space, i.e., their spatial overlap, and (III) the overlap of the wave packets in momentum space, which is referred to as dephasing. Importantly, the semi-classical coherence can be written as a simple product of these contributions
\begin{equation}
    |\chi_{jk}(t)|=c_j(t) c_k(t)
                 e^{-|\mathbf{Q}_{j}(t)-\mathbf{Q}_{k}(t)|^{2}/4\hbar}
                 e^{-|\mathbf{P}_{j}(t)-\mathbf{P}_{k}(t)|^{2}/4\hbar}=(p_jp_k)^{1/2}
                 \mbox{O}_{jk}\mbox{D}_{jk},
\end{equation}
which allows the intuitive interpretation of coherence in terms of populations of the involved electronic states, and positions and momenta of the corresponding nuclear wave packets propagating on those states (see Fig.~4 of the main text).

\subsection{Simulations of strong-field excitation} \label{Sec:SFE}

In order to get insights into the pump step and the resulting initial populations of the silane excited states, we performed simulations by numerically solving the time-dependent Schr\"odinger equation in which the interaction of a neutral silane molecule with the pump laser was treated within the dipole approximation. Starting from the ground state, we perturb the system with a linearly polarized chirped Gaussian electric field
\begin{equation}
\label{eq:field}
    E(t)=E_0 \exp{\left[ -\frac{(t-t_{0})^2}{2 \sigma^2} \right]} \cos{((\omega_0+bt) t)},
\end{equation}
where $t_0$ is the center of the pulse, $\sigma$ is the pulse width, $E_0$ is the field strength, $\omega_0$ is the central frequency, and $b$ denotes the parameter associated with a linear variation of the instantaneous frequency of the pulse in time. The energy levels and the corresponding transition-dipole moments were computed at the EOM-CCSD/aug-cc-pVTZ level of theory. All electronic states lying below 15.3~eV have been taken into account.

\begin{figure}[h!]
\centering
\includegraphics[width=0.7\textwidth]{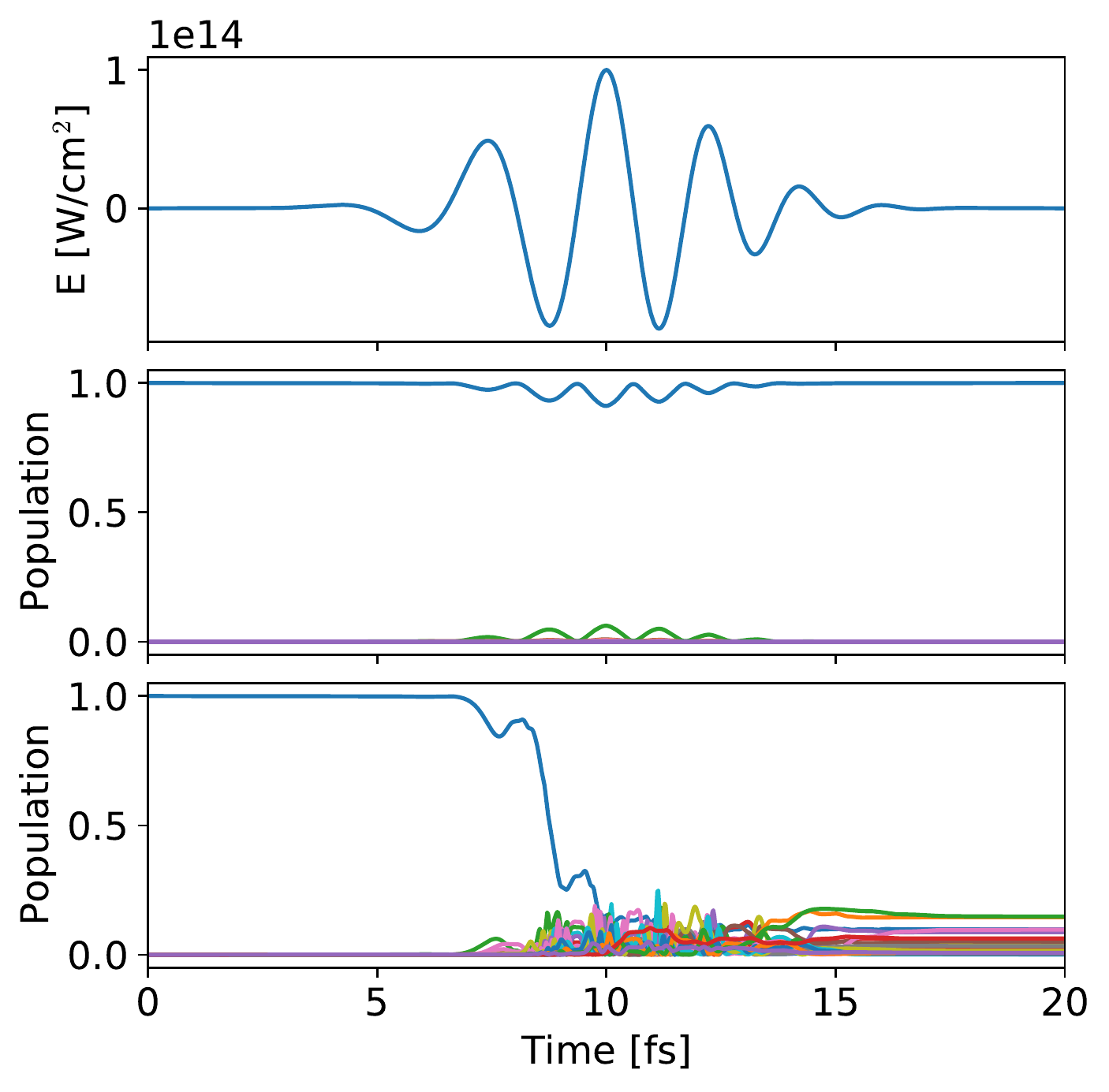}
\caption{\label{pop_transf}\textbf{Population transfer in neutral silane driven by the interaction with the pump pulse.} Top panel: Laser pulse obtained through Eq.~(\ref{eq:field}) using the following parameters: $I_0=c \epsilon_0 E_0^2/2= 10^{14}$~W/cm$^2$, FWHM $=2 \sqrt{2 \ln{2}}\sigma$ = 5.2~fs, $t_0 = 10$~fs, the wavelength spans a range from 950 nm at -3$\sigma$ to 600~nm at +3$\sigma$. Middle panel: Evolution of populations of 60 lowest excited states of a silane molecule driven by such a laser pulse. Only couplings between the ground and the excited states are present while the transitions between the excited states themselves are disabled. Bottom panel: Same as above but all the couplings between the states are taken into account. The energy levels and the corresponding transition dipole moments are computed at the EOM-CCSD/aug-cc-pVTZ level of theory.}
\end{figure}

The evolution of the populations of the excited states driven by a 5.2-fs full-width-at-half-maximum (FWHM $=2 \sqrt{2 \ln{2}}\sigma$) laser pulse with peak intensity of $I_0=c \epsilon_0 E_0^2/2= 10^{14}$ W/cm$^2$ and wavelength spanning the range from 600~nm to 950~nm (at $\pm 3\sigma$) is shown in Figure~\ref{pop_transf}. The middle panel of the figure illustrates the situation when only couplings between the ground and the excited states are present, while the transitions between the excited states themselves are disabled. It is seen that in this case the external field induces the oscillations of the populations between the ground and the excited states during the action of the pulse. However, the system remains almost completely in the ground state as soon as the external field is terminated. The bottom panel of Figure~\ref{pop_transf} depicts the population transfer dynamics for the case when all the couplings between the states are taken into account. As one can see, the external field initiates highly nonlinear population transfer from the ground to all included electronically excited states of the system. The latter indicates the fact that multiphoton excitation plays a central role in the population of excited states of silane by the pump pulse.

Our simulations %by both EOM-CCSD and RT-TDDFT approaches 
demonstrate that the interaction of a neutral silane molecule with the intense pump field leads to a nonlinear population transfer from the ground state to the manifold of electronically excited states. Importantly, we found that the distribution of populations between the excited states is very sensitive to the variations of the laser pulse parameters. However, these calculations had to be carried out in the fixed-nuclei approximation, such that they neglected nuclear motion and the associated non-adiabatic dynamics. These additional effects are addressed in the following subsection.

\subsection{Modeling of multi-cycle pump-pulse effects}

As has been shown in Sec.~\ref{Sec:SFE}, the strong-field excitation is difficult to precisely model because of its high sensitivity to the details of the pump pulse and the fixed-nuclei approximation. In this section we shall, therefore, start from the conclusion of Sec.~\ref{Sec:SFE}, i.e., that all states can, in principle, be significantly populated by the pump pulse but explicitly treat the time evolution of this excitation, in analogy with the recently studied case of strong-field ionization (SFI) \cite{pabst16a}. In that case, due to the highly nonlinear nature of SFI, the majority of the ionization is restricted to the peaks of the electric field in each half-cycle. As a result, a train of wavepackets is launched in the cationic states at a frequency of twice the fundamental driving field. Applying the same temporal concept to strong-field excitation, a train of wavepackets is launched in the valence- and Rydberg-excited states. In contrast to strong-field ionization, however, no photoelectron is emitted (which can lead to entanglement and a loss of coherence) \cite{pabst11a}, therefore, it is only necessary to treat the interference of the molecular wavepackets. 

Let us first consider the electronic wavepacket interference. The phase of the excited electronic states with respect to the ground state at the moment of the creation of a new excited-state wavepacket determines whether the latter will interfere constructively or destructively with the preexisting one. This phase evolves at a frequency that is proportional to the energy difference of the ground and excited state. Since new wavepackets are produced at twice the frequency of the driving field, the population of states with energies close to a $2n\omega$ will interfere constructively and build up over the duration of a multi-cycle pulse while states with energies close to the odd-integer multiples of $\omega$ will experience destructive interference and will therefore be suppressed. 

\begin{figure}[h!]
\centering
\includegraphics[width=\textwidth]{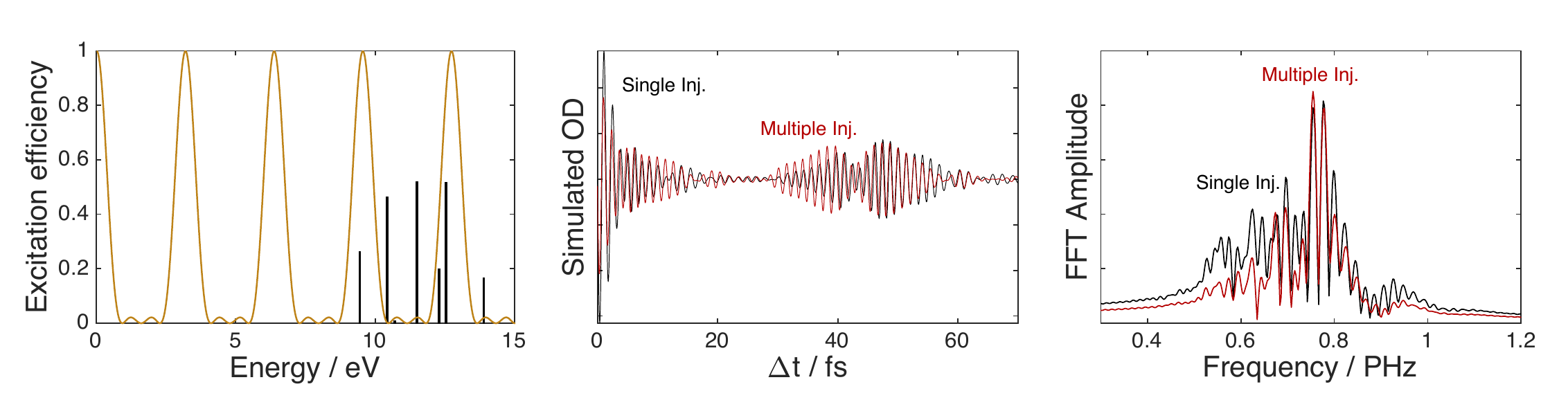} \label{MulticycleFig}
\caption{\textbf{Inter-half-cycle interference of strong-field-prepared wavepackets} Left: Energy dependence of the excitation efficiency of an electronic state due to the interference of wavepackets created in four consecutive half-cycles of a 780~nm pump pulse. The black lines indicate the energy and transition-dipole moment of the electronically excited states of silane. Middle: Comparison between the MCTDH simulated quantum beats of a simulation with a single wavepacket initialization (black, same as Fig.~S8) and a simulation in which four identical wavepackets were launched in each electronic state with delays of 1.3~fs (red). Right: FFT of the middle panel.}
\end{figure}

Our D-scan reconstruction of the pump pulse shows that up to four half-cycles can contribute to the strong-field excitation with relative intensities of 2:3:3:2. Hence, the energy dependence of the pump-pulse's excitation efficiency, due to its multi-cycle nature, can be found by calculating the square amplitude of the sum of four cosine oscillations at a relative delay of 1.3~fs (corresponding to the center wavelength of 780~nm) as a function of the oscillating frequency. This is equivalent to finding the squared Fourier transform of a sequence of four delta functions spaced by 1.3~fs (equal to the product of a rectangular function and a Dirac comb in the time domain). By making use of the convolution theorem we find that the Fourier transform of such a function is the convolution of a sinc function and a Dirac comb in the frequency domain. Indeed, Fig.~\ref{MulticycleFig} which present the excitation efficiency of the pump pulse, display a comb of sinc-squared functions with three nodes (one less than the number of pulses) between subsequent maxima. The proximity of the A, B, and C states to the peaks of excitation efficiency further explains why these coherences survive the multi-cycle excitation, whereas other possible coherences do not. This result additionally supports our experiment-based assignment of the attosecond electron dynamics to the B-C and A-C coherences.

Vibrational dynamics have the potential, through the displacement of the nuclear wavepacket away from the Frank-Condon region between half-cycles, to reduce the coherence of the electronic wavepackets at the moment of their creation and therefore halt the constructive interference that leads to the energetically selective excitation. In the case of silane, however, this effect is relatively small because the half cycle of the pump pulse is short in comparison with the vibrational period. It is estimated that such effects would become observable with longer-wavelength pump pulses of at least 4~\(\mu\)m. To confirm the lack of effect of the nuclear dynamics on the coherence buildup, MCTDH calculations were run in which multiple wavepackets were consecutively launched from the Frank-Condon region into all contained electronic states. These showed that the absolute value of the initial coherence is only reduced by 20\% at the time of the following population injection, enabling a buildup of the electronic coherence over multiple cycles of the pump pulse. Furthermore, the multiple injection MCTDH simulations showed an increase of the A (+25\%), B (+50\%), and C (+50\%) states with no significant change to the temporal evolution of their quantum beats (which can be seen in the middle panel of Fig.~\ref{MulticycleFig}). In fact, since the multi-injection simulation lowers the relative population of the higher-energy valence states 3, 4 and 5 by 20\%, 70\% and 99\% respectively, the total quantum beat of the system becomes even more heavily dominated by the 0.7-0.8~PHz signal of the A-B and A-C coherences (see right-hand panel of Fig.~\ref{MulticycleFig}), leading to a simulated OD signal that resembles the experimentally measured quantum beats even more closely. 
We therefore conclude that, when treated with the strong-field-excitation model described here, the few-, but still multi-cycle nature of our pump-pulse has the potential to energetically selecting well-defined subsets of electronic coherences. The $2n\omega$ spacing of the energy windows of excitation is accordingly reflected in the higher likelihood of observing a quantum beat with an energy in the vicinity of $2\omega$. These results imply that the lack of strong signals at 0.65~PHz and 0.58~PHz in the experimental results could not only be due to very fast non-adiabatic depopulation of these states but also due to the energetically selective preparation of electronic coherences. 

\clearpage

%%-------------------- style1
\bibliography{attoH2On,attobib,biblio_SOM_Theory_Silane, extra}
\bibliographystyle{Science}

%-------------------- style2